\documentclass[traditabstract]{aa}
\usepackage{txfonts}
\usepackage{natbib}
\usepackage{graphicx}
\usepackage{rotating}
\usepackage{aalongtable}
\usepackage{float}
\newcommand{\Msun}{M$_{\odot}$}
\newcommand{\eso}{ESO-H$\alpha$ 574 }
\newcommand{\para}{Par-Lup 3-4 }
\newcommand{\xsh}{X-Shooter }
\newcommand{\Lsun}{L$_{\odot}$}

\newcommand{\Rsun}{R$_{\odot}$}

\newcommand{\Ha}{H$\alpha$}

\newcommand{\km}{km~s$^{-1}$}

\begin{document}

\title{ESO-H$\alpha$~574 and Par-Lup3-4 Jets: Exploring the spectral, kinematical and physical properties\thanks{Based on Observations collected with X-shooter and UVES at the Very Large Telescope on Cerro Paranal (Chile), operated by the European Southern Observatory (ESO). Program ID's: 085.C-0238(A) and 078.C-0429(A).}}

\author{Whelan, E.T. \inst{1}
  \and 
 Bonito, R \inst{2,3} 
  \and 
 Antoniucci, S \inst{4}
 \and 
 Alcal{\'a}, J.M. \inst{5 }
 \and
 Giannini, T. \inst{4}
 \and
 Nisini, B. \inst{4}
 \and
 Bacciotti, F.  \inst{6}
 \and
  Podio, L. \inst{6, 7}
 \and
 Stelzer, B. \inst{3}
 \and
 Comer{\'o}n, F. \inst{8}}

\institute{Institut f{\"u}r Astronomie und Astrophysik, Kepler Center for Astro and Particle Physics, Eberhard Karls Universit{\"a}t,  72076 T{\"u}bingen, Germany  \and  Universit{\`a} di Palermo, P.zza del Parlamento 1, 90134 Palermo, Italy \and INAF-Osservatorio Astronomico di Palermo, Piazza del Parlamento 1, 90134 Palermo, Italy \and INAF-Osservatorio Astronomico di Roma, via Frascati 33, 00040 Monte Porzio, Italy \and INAF-Osservatorio Astronomico di Capodimonte, via Moiariello, 16, 80131, Napoli, Italy \and INAF-Osservatorio Astrofisico di Arcetri, Largo E. Fermi 5, 50125 Firenze, Italy \and  UJF-Grenoble 1 / CNRS-INSU, Institut de Planetologie et d'Astrophysique de Grenoble (IPAG) UMR 5274, Grenoble, 38041,
France \and ESO,  Alonso de C{\'o}rdova 3107, Castilla 19001, Santiago 19, Chile}

\titlerunning{X-Shooter observations of the \eso and \para Jets} 
\date{}

\abstract {


In this paper a comprehensive analysis of VLT / X-Shooter observations of two jet systems, namely ESO-H$\alpha$ 574 a K8 classical T Tauri star and
Par-Lup 3-4 a very low mass (0.13~\Msun) M5 star, is presented. Both stars are known to have near-edge on accretion disks. A summary of these first X-shooter observations of jets was given in a 2011 letter. The new results outlined here include flux tables of identified emission lines, information on the morphology, kinematics and physical conditions of both jets and, updated estimates of $\dot{M}_{out}$ / $\dot{M}_{acc}$. Asymmetries in the \eso flow are investigated while the \para jet is much more symmetric. The density, temperature, and therefore origin of the gas traced by the Balmer lines are investigated from the Balmer decrements and results suggest an origin in a jet for \eso while for \para the temperature and density are consistent with an accretion flow. $\dot{M}_{acc}$ is estimated from the luminosity of various accretion tracers. For both targets, new luminosity relationships and a re-evaluation of the effect of reddening and grey extinction (due to the edge-on disks) allows for substantial improvements on previous estimates of $\dot{M}_{acc}$. It is found that log($\dot{M}_{acc}$) = -9.15 $\pm$ 0.45~\Msun yr$^{-1}$ and -9.30 $\pm$ 0.27~\Msun yr$^{-1}$ for \eso and \para respectively. Additionally, the physical conditions in the jets (electron density, electron temperature, and ionisation) are probed using various line ratios and compared with previous determinations from iron lines. The results are combined with the luminosity of the [SII]$\lambda$6731 line to derive $\dot{M}_{out}$ through a calculation of the gas emissivity based on a 5-level atom model. As this method for deriving $\dot{M}_{out}$ comes from an exact calculation based on the jet parameters (measured directly from the spectra) rather than as was done previously from an approximate formula based on the value of the critical density at an assumed unknown temperature, values of $\dot{M}_{out}$ are far more accurate. Overall the accuracy of earlier measurements of $\dot{M}_{out}$ / $\dot{M}_{acc}$ is refined and $\dot{M}_{out}$ / $\dot{M}_{acc}$ = 0.5 (+1.0)(-0.2) and 0.3 (+0.6)(-0.1) for the \eso red and blue jets, respectively, and 0.05 (+0.10)(-0.02) for both the \para red and blue jets.  While the value for the total (two-sided) $\dot{M}_{out}$ / $\dot{M}_{acc}$ in \eso lies outside the range predicted by magneto-centrifugal jet launching models, the errors are large and the effects of veiling and scattering on extinction measurements, and therefore the estimate of $\dot{M}_{acc}$, should also be considered. ESO-H$\alpha$ 574 is an excellent case study for understanding the impact of an edge-on accretion disk on the observed stellar emission. The improvements in the derivation of $\dot{M}_{out}$ / $\dot{M}_{acc}$ means that this ratio for \para now lies within the range predicted by leading models, as compared to earlier measurements for very low mass stars. \para is one of a small number of brown dwarfs and very low mass stars which launch jets. Therefore, this result is important in the context of understanding how $\dot{M}_{out}$ / $\dot{M}_{acc}$ and, thus, jet launching mechanisms for the lowest mass jet driving sources, compare to the case of the well-studied low mass stars.}


\keywords{interstellar medium: jets and outflows --
 stars: pre-main-sequence}
\maketitle

\section{Introduction}

 The mass outflow phase is a key stage in the star formation process. Protostellar jets (fast, collimated outflows) are an important manifestation of the outflow phenomenon and it is generally accepted that they are strongly connected with accretion \citep{Cabrit07}. The importance of these jets lies in the fact that they are the likely mechanism by which angular momentum is removed from the star-disk system \citep{Coffey04}. While jets from low mass young stellar objects (YSOs) have been studied now for several decades, more recently it has been found that very low mass stars (VLMS) and brown dwarfs (BDs) also drive jets during their formation \citep{Joergens13}. The jets of VLMSs and BDs have many similarities with jets of low mass YSOs. For example, they have been found to be collimated, episodic, and asymmetric \citep{Whelan09, Whelan12}, they can have multi-component velocity profiles \citep{Whelan09}, and they are associated with a molecular counterpart \citep{Monin13}. Therefore, it is possible that the mechanisms responsible for the launching and collimation of protostellar jets also operate down to substellar masses \citep{Whelan09}.

Although a magneto-centrifugal jet launching model is favoured for the production of jets, the precise scenario is still debated \citep{Ferreira13, Frank14} and, thus observational constraints are currently needed. Protostellar jets are characterised by an abundance of shock excited emission lines and analysis of these emission regions, using both spectroscopy and imaging, offers a wealth of information pertinent to jet launching models. Of particular value are spectroscopic observations which simultaneously cover different wavelength regimes. Such observations allow outflow and accretion properties to be investigated from the same dataset and over a significant range in wavelength, leading, for example, to more accurate estimates of the ratio of mass outflow to accretion ($\dot{M}_{out}$ / $\dot{M}_{acc}$).  For instance, studies have shown that the optimum way to measure $\dot{M}_{acc}$ is by using several different accretion indicators, found in different wavelength regimes \citep{Rigliaco11, Rigliaco12}. In this way any spread in $\dot{M}_{acc}$ due to different indicators probing different regimes of accretion or having varying wind/jet contributions can be overcome. Magneto-centrifugal jet launching models place an upper limit (per jet) of $\sim$ 0.3 on $\dot{M}_{out}$ / $\dot{M}_{acc}$  \citep{Ferreira06, Cabrit09}. Therefore it is important to measure this ratio not only in low mass YSOs but also in VLMSs and BDs. $\dot{M}_{out}$ / $\dot{M}_{acc}$ has been constrained at 1~$\%$ to 10~$\%$ for T Tauri stars (class II low mass YSOs; TTSs; \cite{ Hartigan95, Melnikov09, Agra09}). Initial attempts at estimating  $\dot{M}_{out}$ / $\dot{M}_{acc}$ in VLMSs / BDs produced results which suggested that this ratio is higher in BDs than in TTSs and at most to be comparable \citep{Comeron03, Whelan09}. More recent studies of the VLMS ISO 143 and the BD FU Tau A, produced ratios which are more in line with studies of TTS \citep{Joergens12b, Stelzer13} . Thus much work is still needed to constrain $\dot{M}_{out}$ / $\dot{M}_{acc}$ at the lowest masses. This work will involve overcoming various observational uncertainties which are outlined in the above papers and addressed as part of this work.

X-Shooter, one of the newest instruments on the European Southern Observatory's (ESO) Very Large Telescope (VLT), provides contemporaneous spectra in the ultraviolet (UVB), visible (VIS), and near-infrared (NIR) regimes, with a total coverage of $\sim$ 300~nm to 2500~nm. With the aim of exploring the advantages of \xsh for the study of outflow and accretion activity in YSOs / BDs, we conducted in 2010 a pilot \xsh study of two YSOs. The targets were the classical T Tauri star (CTTS) ESO-H$\alpha$ 574 and the VLMS Par-Lup3-4 (see Section 2). As these were the first \xsh observations of YSO jets the initial results and spectra were published in a letter \citep{Bacciotti11}. In addition, the numerous [Fe II] and [Fe III] detected lines have been discussed in Giannini et al. (2013),
where the potential of UV-VIS-NIR Fe line diagnostics in deriving the jet physical parameters is demonstrated. The [Fe II] line analysis revealed that the ESO-H$\alpha$ 574 jet is, on average, colder, less dense, and more ionised than the Par-Lup 3-4 jet. The physical conditions derived from the iron lines were also compared with shock models which pointed to the ESO-Hα 574 shock likely being faster and more energetic than the Par-Lup 3-4 shock. 
In \cite{Bacciotti11}, outflow and accretion
tracers present in the spectra
were examined, the effect of extinction by an edge-on disk
was discussed, and first estimates of $\dot{M}_{acc}$
and the jet parameters n$_{e}$ and $\dot{M}_{out}$ were given.
However, this analysis was tentative, and is now expanded
and improved on here with new up-to-date procedures. 
The study of \para was relevant to the question of the applicability of jet launching models at the lowest masses and one of the particular aims of the work presented here was to overcome difficulties with previous estimates of  $\dot{M}_{out}$ / $\dot{M}_{acc}$ in Par-Lup3-4.

Firstly, tables of all the emission lines detected in the spectra of the two YSOs and their measured fluxes are presented. The goal here is to provide an important reference for future observational and computational studies of jets. Section 4.1 describes the line identification process and the tables of identified lines can be found in Appendices A1 and A2.  Secondly, the kinematics and morphology of the two jets are discussed in greater detail than in \cite{Bacciotti11} (Sections 4.2 and 4.3).  For ESO-H$\alpha$ 574, high spectral resolution UVES spectra taken from the ESO archive are included in order to improve the kinematical analysis. Since \cite{Bacciotti11}, several \xsh studies of accretion in YSOs have been conducted with participation from members of our team \citep{Rigliaco12, alcala13, Manara13}. This experience has allowed us to refine our methods for estimating $\dot{M}_{acc}$. Hence, thirdly an updated analysis of $\dot{M}_{acc}$ in both targets is given, including a more detailed investigation of how the extinction of both sources can be evaluated and how it effects $\dot{M}_{acc}$ estimates (Sections 4.4, 4.5). As both targets have edge-on disks, the obscuration affects of the disks are particularly relevant to our study of $\dot{M}_{acc}$. Furthermore, an improved approach for measuring $\dot{M}_{out}$ is presented along with a more accurate determination of  $\dot{M}_{out}$ /  $\dot{M}_{acc}$ for both sources. Finally, the origin of the permitted emission in both sources is explored by examining their Balmer decrements (see Section 4.6).

\begin{figure}
   \includegraphics[width=11cm, trim= 3cm 0cm 0cm 0cm, clip=true]{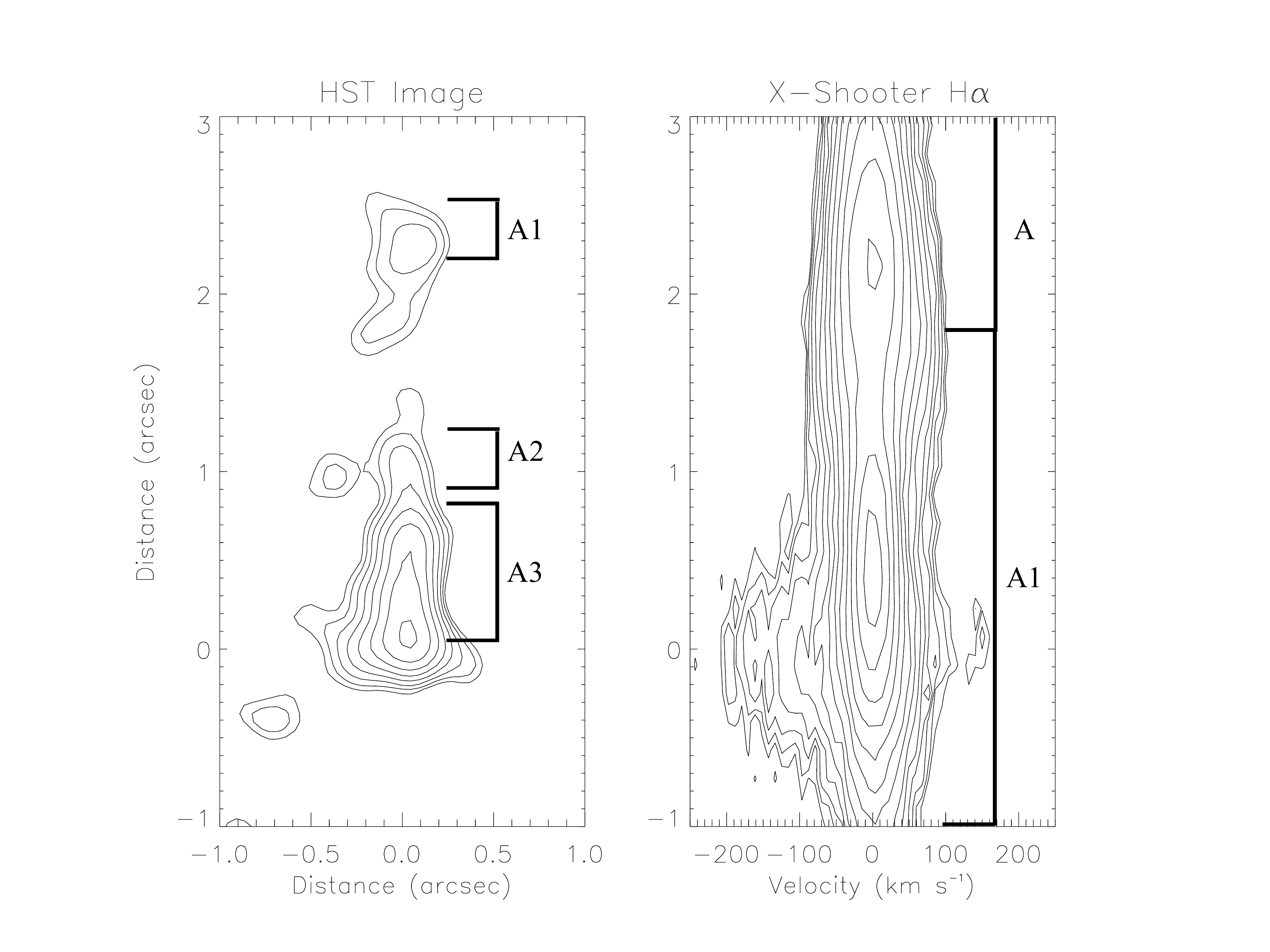}
     \caption{Different notation systems for the knots of \eso found in the literature. Left: HST WFPC images of \eso published in \cite{Robberto12}. For the purpose of this figure we obtained the reduced image from the HST archive (http://archive.stsci.edu/hst/). Right: Position-velocity diagram of the \eso jet in H$\alpha$ made from our \xsh data and first published in \cite{Bacciotti11}. In this paper we follow the notation of \cite{Bacciotti11}. }
  \label{comp_eso}     
\end{figure}

 \section{Source properties}

 \subsection{\eso}
 
ESO-H$\alpha$ 574 (11$^{h}$16$^{m}$03$^{s}$7, -76$^{\circ}$24$^{\arcmin}$53$^{\arcsec}$) located in the Chamaeleon I star-forming region (d = 160 $\pm$ 17~pc, Wichmann et al. 1998) is a low luminosity source with a derived spectral type of K8 \citep{Comeron04, Luhman07}. Previous to this study it was known that \eso powered a jet bright in many lines (HH 872 of total length 3150~AU) \citep{Comeron06}. The position angle (PA) of the jet is $\sim$ 45$^{\circ}$ with up to 6 knots detected in the blue-shifted lobe (knots A to D) and a single knot, knot E in the red-shifted lobe \citep{Bacciotti11, Robberto12}. \cite{Robberto12} used the Wide Field Planetary Camera on the HST to separate knot A into three separate knots which they label A1, A2, and A3. In the \xsh observations A2 and A3 are detected as one single knot labelled A1 by \citep{Bacciotti11} and A1 from Robberto et al. (2012) refers to knot A from Bacciotti et al. 2011. Also refer to Figure \ref{comp_eso} for further clarification on the naming of the knots. 

The edge-on nature of the \eso accretion disk was first reported in \cite{Comeron04}. The prediction of an edge-on disk for this star was based on the observation that the accretion signatures of \eso were suppressed with respect to the outflow signatures suggestive of obscuration of the accretion zone by the disk.  In addition, its bolometric luminosity is only 0.0034 \Lsun\ \citep{Luhman07}, which, when compared with other typical T Tauri stars of the same spectral type (K8) in Chamaeleon I, makes this star under-luminous by a factor of about 150. This under-luminosity can be caused by an edge-on accretion disk. Finally the low radial velocities measured in the jet, as reported in \cite{Bacciotti11}, are typical of a star with an edge-on disk and therefore a jet which lies in the plane of the sky. The properties of ESO-HA 574 are very similar to TWA30 which is also a strong candidate for having an edge-on accretion disk. \cite{Looper10a} discuss how the low radial velocities of the numerous forbidden emission lines (FELs) detected in the spectrum of TWA30, is evidence of an edge-on disk.

 \subsection{\para}

Par-Lup3-4 (16$^{h}$08$^{m}$51$^{s}$44, -39${\circ}$05$^{\arcmin}$30.5$^{\arcsec}$) found in the Lupus III cloud (d = 200 $\pm$ 40~pc, Comeron et al. 2003), is a VLM object driving a small-scale jet at a PA of $\sim$ 130$^{\circ}$ (HH 600) \citep{Comeron11}.  Its spectral type and mass have been estimated at M5 and 0.13~\Msun\ respectively \citep{Comeron03, alcala13}. The jet from \para was first reported by \cite{Comeron05}. It was discovered using both spectroscopy and imaging and it is one of the lowest mass objects for which a jet has been directly imaged \citep{Whelan12}. The source is under-luminous by $\approx$ 4 mag with respect to other M5 young stars in Lupus III  \citep{Comeron03, alcala13}. \cite{Nuria10} model the SED (from the optical to submillimetre) of \para and conclude that it also has an edge-on disk with an estimated inclination of 81$^{\circ}$ $\pm$ 6$^{\circ}$. The fact that \para has stronger accretion tracers than \eso \citep{Bacciotti11} indicates that it may have a near edge-on disk, which is slightly more inclined than the ESO-H$\alpha$ 574 disk.

\section{Observations and data reduction}

The \xsh observations presented in this paper were conducted on the VLT as part of the INAF \xsh GTO programme ``An X-Shooter survey in star formation regions: low and sub-stellar mass objects'' (085.C-0238(A), see \cite{alcala11}). Both \eso and \para were observed on April 7 2010. {\rm For details of the observations and of the data reduction performed with the \xsh pipeline please refer to \cite{Bacciotti11}}. Flux calibration was achieved using context {\em LONG} within {\em MIDAS}. 
For this purpose, a response function was derived by interpolating the 
counts/standard-flux ratio of the flux standards, observed the same night as 
the objects, with a third-order spline function, after airmass correction. 
The response function was then applied to the 1D spectra. Following the independent flux calibration of the \xsh arms,  the internal
consistency of the calibration was checked by plotting together the three spectra extracted
from the source position and visually examining the superposition of overlapping
spectral regions at the edge of each arm. For the observations of ESO-H$\alpha$ 574, the UVB and VIS arms were found to be very well
aligned, while the NIR arm presented a shift with respect to the
VIS spectrum of $\sim$ 25$\%$ lower. This was corrected for by scaling the NIR spectrum to the VIS
continuum level. In the case of the \para observations the alignment of the three arms was found to be very good and no re-scaling was necessary.

Previously unpublished UVES observations of \eso (078.C-0429(A)) are also included in this paper. See \cite{Dekker00} for a description of this instrument. The UVES spectra were obtained on December 12 2006 and the exposure time was 1000~s. A slit of width 1~\arcsec\ was again placed along the known jet PA. The resultant spectral resolution was $\sim$ 40 000 for the UVES data and the pixel scale is 0\farcs182. The wavelength range of the UVES observation was approximately 4750~\AA\ to 6850~\AA\ but poor observing conditions  meant that the jet is only detected in \Ha, [O I]$\lambda\lambda$6300, 6363 and [S II]$\lambda\lambda$6717, 6731 with a good signal-to-noise ratio ( $\sim$ 10). The seeing averaged 1\farcs5 during the observation. The UVES spectra were reduced using standard routines for bias subtraction, flat fielding, and wavelength calibration provided by IRAF. 
 
 \begin{figure*}
   \includegraphics[width=23cm, trim= 0cm 0cm 0cm 0cm, clip=true, angle=-90]{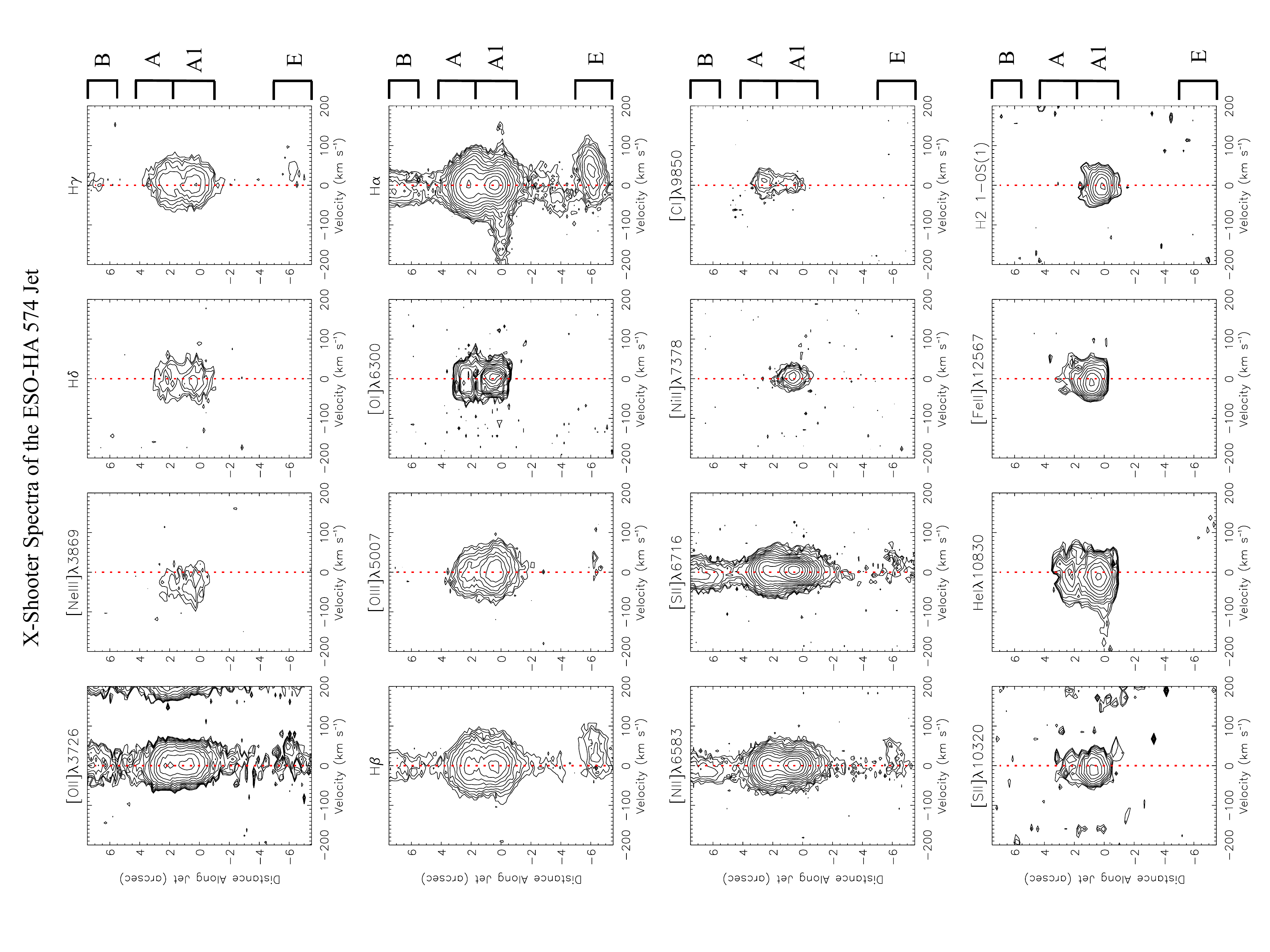}
     \caption{Position velocity diagrams of the \eso jet in some of the most interesting jet lines. Contours begin at 0.16 $\times$ 10$^{-17}$ ergs/s/cm$^{2}$ and increase logarithmically. Velocities are systemic. The positions and size of each knot are marked. The notation system of \cite{Bacciotti11} is used. See Figure 1 for the notation system of \cite{Robberto12}.}
  \label{mix}     
\end{figure*}

\section{Results}

\subsection{Line identification / flux measurement and summary of spectral characteristics}

By placing the \xsh slit along the PAs of both jets and using the nodding mode the jet emission is traced to 8~\arcsec\ on either side of the driving source. For \eso this includes knots A1 to B in the blue lobe and knot E in the red. The same labels for the knots as adopted by \cite{Bacciotti11} are used here, and A2 and A3 from the notation of  \cite{Robberto12} and not resolved here. The full known extent of the \para jet is covered by the \xsh slit. Emission lines were identified using the Atomic Line List database
\footnote{(http://www.pa.uky.edu/~peter/atomic/)}. For the identification, we considered
a wavelength uncertainty of about 0.5 \AA. Nebular lines from abundant
species having excitation energies below 40 000 cm$^{-1}$ were firstly searched for, and only when identification with these criteria was not successful were permitted lines from neutral or single ionised atoms also considered. Line fluxes were computed through the Gaussian fitting of the line profiles after subtraction of the local continuum. 
Gaussian fitting was done using the IRAF task {\it splot}. Absolute flux errors
were computed from the root mean square (r.m.s) noise (measured in a portion of the spectrum adjacent
to the line) multiplied by the spectral resolution element at the considered wavelength.

All the lines identified in the spectra of both targets along with the observed wavelengths and the lines fluxes 
are listed in Tables A1 and A2. In the case of \eso (Table A1) the line fluxes are for the region of knot A1. Where emission lines are also detected in the other knots these fluxes are also given. The extraction regions for the knots A1, A, B, and E are -1\farcs0 to 1\farcs8, 1\farcs8 to 4\farcs2, 5\farcs6 to 7\farcs8, and -7\farcs5 to -5\farcs2, respectively (also see Figure \ref{mix}). For \para (Table A2) the fluxes are measured from the spectrum extracted at the source position (-0\farcs5 to 0\farcs5). A small number of lines are also seen in the Par-Lup3-4 jet and for these lines the fluxes in the blue and red-shifted jets are measured over the regions 0\farcs5 to 2\farcs4 and -0\farcs5 to -3\farcs3, respectively. The line fluxes are not corrected for extinction and a discussion on estimating the extinction of \eso and \para is given in Section 4.4. 

From studying Table A1 it is clear that \eso has a particularly rich spectrum.  
However, it is the abundance of jet lines that makes its spectrum so rich while the accretion tracers are not so common and those that are seen are weak.
The \eso jet is detected in many tens of emission lines, in species of hydrogen, oxygen, iron, sulphur, nitrogen, helium, and calcium for example, and in molecular lines of H$_{2}$. In Figure 2 the position velocity (PV) diagrams of a selection of lines tracing the \eso jet are presented. Numerous Balmer lines, although weak, are detected, while from the Paschen series only Pa$\gamma$ and Pa$\beta$ are observed and no Brackett lines are present \citep{Folha01}. Also note that the Ca II triplet is much fainter than would be expected for  a CTTS which is driving such a powerful jet. Due to the weakness and absence of key accretion indicators, \eso cannot be said to have a spectrum which is typical of a CTTS. However, its spectrum could be said to be characteristic of a CTTS with a disk with an inclination close to 90$^{\circ}$. Note the similarity of the ESO-HA 574 spectrum to that of TWA30, a YSO with an edge-on disk investigated by \cite{Looper10a}. While accreting YSOs with edge-on disks have been seen before, the effect of the disk obscuration on the accretion indicators is not often so stark. Thus these observations of \eso are a good point of reference for future observations of similar YSOs. \para has a spectrum which is typical of a young VLMS, in that all the usual accretion tracers are found while the jet is detected in only a handful of lines, e.g. [OI]$\lambda$6300, [SII]$\lambda$6731 \citep{Whelan09}. Finally note that both sources show an abundance of Fe emission lines \citep{teresa13}. The full \xsh spectra of both targets are shown in \cite{Bacciotti11}.

\subsection{Kinematics and morphology of the \eso jet}

Any kinematical study of the \eso jet based on spectroscopic data is restricted by the edge-on nature of the \eso disk and the resultant low inclination angle of the jet with respect to the plane of the sky.  Outflows in the plane of the sky are characterised by low systemic radial velocities. This is especially true in the case of \eso, where the bulk of the jet emission lies along the 0 \km\ line  (Figure \ref{mix}). The low jet inclination angle was confirmed in \cite{Bacciotti11} where spectroscopic observations of \eso were compared to estimate the proper motions and thus tangential velocities and inclination angles of the knots. The effect of the low radial velocities on the kinematical analysis is compounded by the intermediate resolution of X-Shooter and for this reason archived UVES observations are also introduced here. Overall our  discussion of the kinematics of the \eso jet is confined to a discussion of asymmetries and of the shape of the \Ha\ line profile.


\begin{figure}[h!]
   \includegraphics[width=13cm, angle=-90]{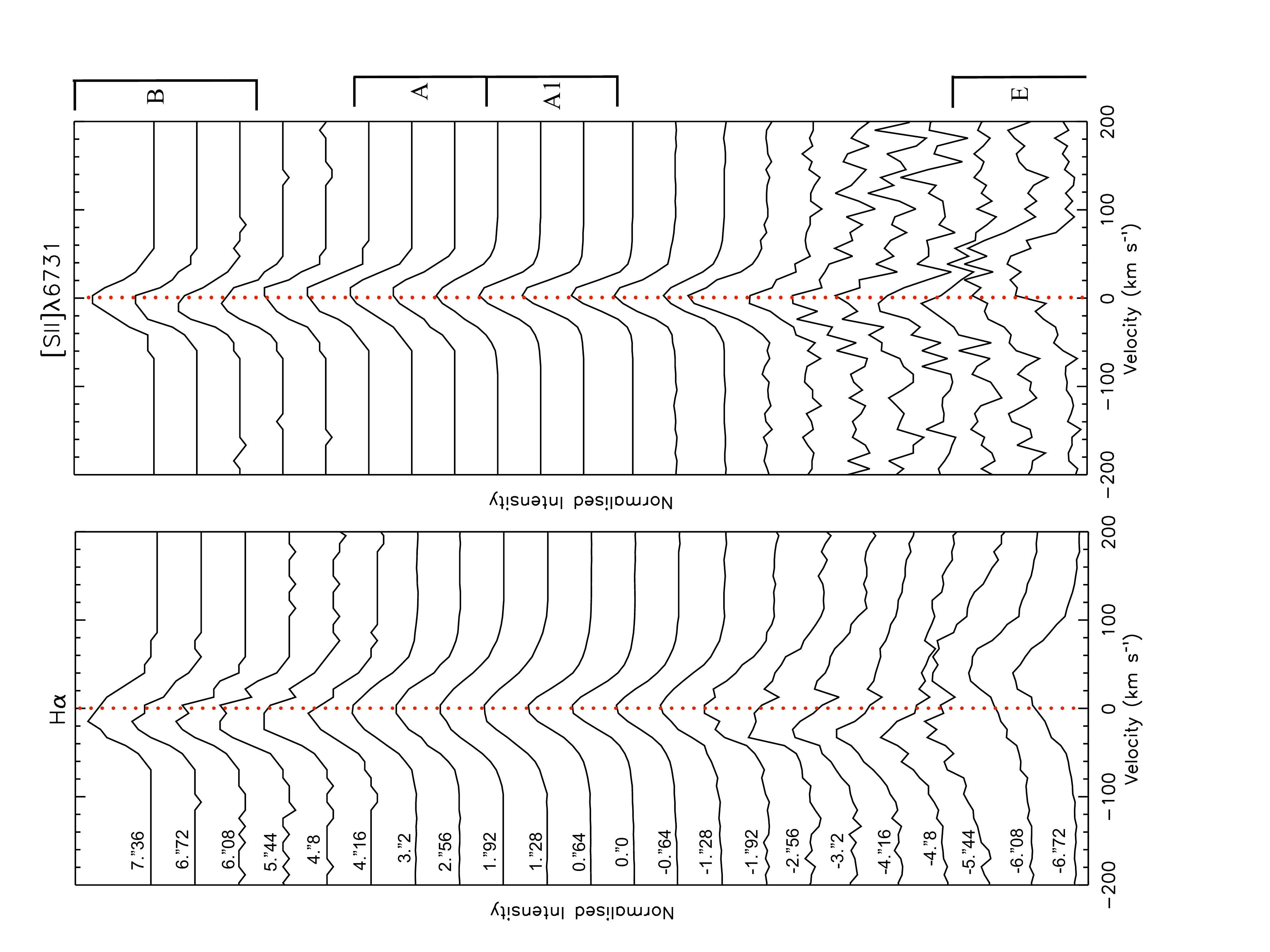}
     \caption{Line profiles extracted in \Ha\ and [SII]$\lambda$6731 from the \xsh spectra at various points along the length of the \eso jet. While knot E at $\sim$ - 6.5~\arcsec\ is noisier than the blue-shifted part of the jet it has a higher velocity.}
 \label{velocity_prof}        
\end{figure}


\begin{figure}
\centering
   \includegraphics[width=14.5cm, angle=-90, trim= 0cm 2.5cm 0cm 0.0cm, clip=true]{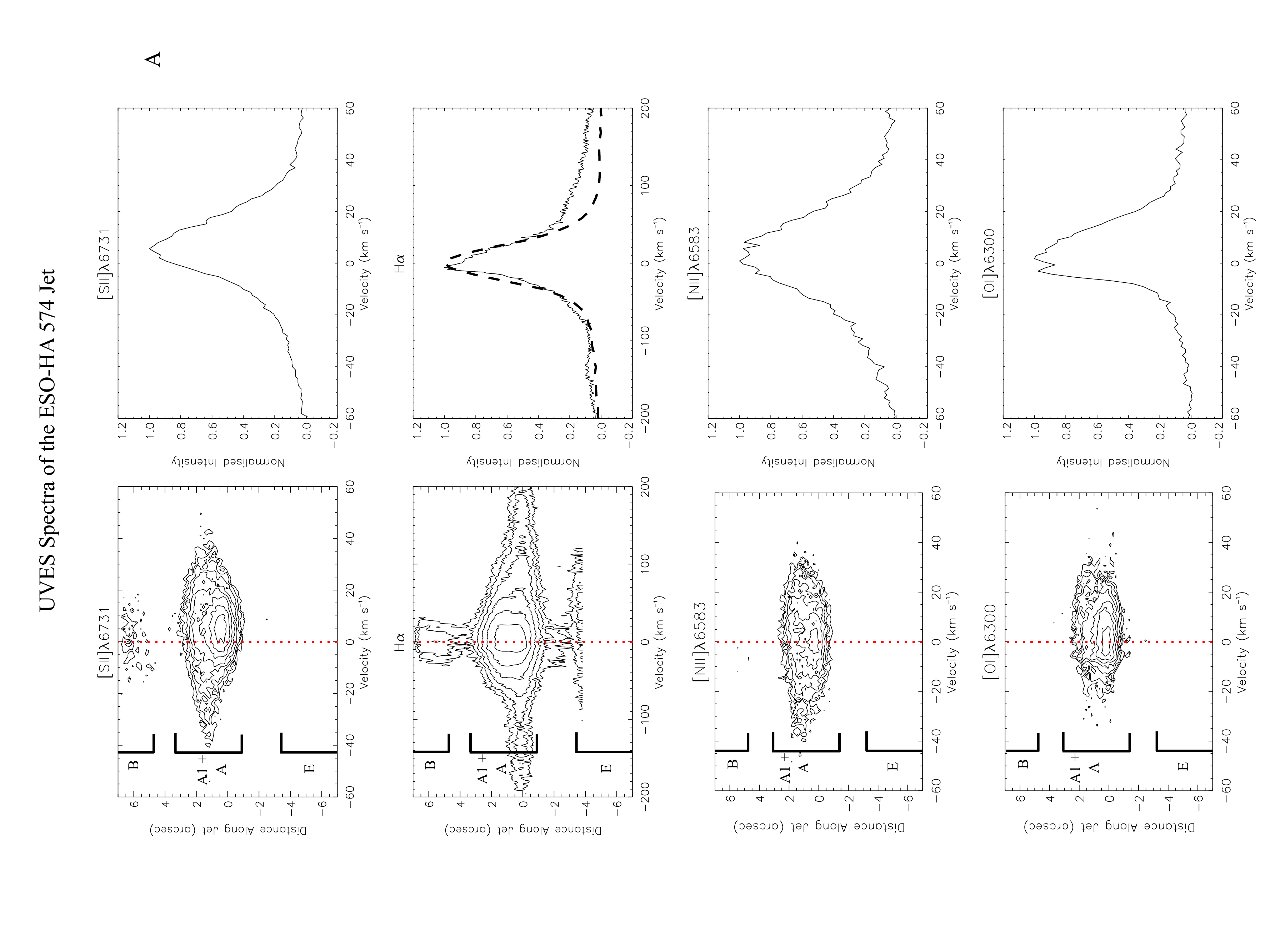}
     \caption{Position velocity diagrams and line profiles of various lines in the \eso jet made from UVES spectra taken in 2006.  All velocities are systemic. Knots A1, A, and some of knots B and E are included in the data. Knots A1 and A are not resolved. Note that due to a difference of $\sim$ 3.3~yrs between this data and the X-Shooter data the knots here are closer to the source positions than in Figure 2. Using the tangential velocities given in Table 1, we calculate that knots A, B, and E will have shifted by 0\farcs6, 0\farcs95, and 1\farcs4, respectively.The [S II]$\lambda$6731 and [N II]$\lambda$6583 lines regions have a blue-shifted wing which increases in velocity with distance from the source. In contrast to the \Ha\ line region in the \xsh spectra the \Ha\ line here has a more pronounced red-wing. For comparison the \xsh \Ha\ line is over-plotted on the UVES profile as a dashed line. The [O I]$\lambda$6300 line region is slightly distorted from the subtraction of the [O I] sky line.}
 \label{uves}        
\end{figure}

\subsubsection{Asymmetries in the jet} All four knots are only seen in a small fraction of the total number of lines which trace the jet, with the H$\beta$, H$\alpha$, [NII]$\lambda$6583, and [SII]$\lambda$6716, 6731 lines providing the most kinematical information (refer to Figure \ref{mix}). 
While knots A1, A, and B lie close to 0~\km\ there is an increase in radial velocity towards knot E of $\sim$ 40~\km. This is highlighted in Figures \ref{mix} and \ref{velocity_prof}.  In Figure \ref{velocity_prof} the extracted X-Shooter line profiles for the \Ha\ and [SII]$\lambda$6731 lines are shown. The difference in velocity between the knots in the blue-lobe and knot E shows that the jet is asymmetric in velocity. The jet spectral profile is extracted at fixed points along the jet by spatially integrating the lines over the seeing full width half maximum (FWHM) of $\sim$ 8 pixels. Gaussian fitting is then used to record the centroid velocity for each of the extracted profiles. In Figure \ref{uves}, PV diagrams of the [S II]$\lambda$6731, H$\alpha$, [N II]$\lambda$6583, and [O II]$\lambda$6300 emission line regions made from the UVES observations are displayed alongside extracted spectral lines profiles. Knots A1 and A are included in the observation however knots B and E are at the edge of the spatial coverage of the spectrum. Some noisy emission from B can be seen in the [SII]$\lambda$6731 PV diagram. While the bright knot of emission at the source position is centred at 0~\km, note the presence of a blue-wing in the [S II]$\lambda$6731 and [NII]$\lambda$6583 lines which extends to $\sim$ 40~\km. This is evidence of some acceleration in the jet material with distance from the star, as seen for the red-lobe in the \xsh spectrum. The blue wing is also visible in the line profiles. Again the spectral profiles are extracted by summing over the seeing FWHM and were taken from the region at $\sim$ 1\farcs5 from the source. The blue-wing is absent in the [O I]$\lambda$6300 line but this is due to an imperfect subtraction of the strong [O I]$\lambda$6300 sky line.

The spacing and number of the knots is also asymmetric in that there are several close knots in the blue-lobe which start at the source, while in the red lobe knot E at $\sim$ 6\arcsec\ is the first distinct emission knot. 
For protostellar jets, the blue-shifted jet is commonly observed to be the dominant lobe as the red-shifted flow is obscured by the disk within a few arcseconds of the driving source. However, in cases where the jet has a low inclination angle the obscuration is minimised and the red-shifted flow can be traced back to the star. Therefore, for \eso it seems likely the observed asymmetry is as a result of the kinematics of the jet and points to differences in the rate at which the jet ejects material into both lobes and to differences in the medium into which the lobes are propagating. Variable ejection is commonly seen in jets and can be explained in terms of a pulsed jet model. 
\cite{Bonito10a} put forward a model of a jet ejected with variable velocity. The jet propagates into an inhomogeneous ambient medium with the inhomogeneity resulting from the knots ejected by the source interacting with the medium itself.
A mutual interaction of knots arises from the random velocity of ejection of material at different epochs together with the strongly inhomogeneous medium into which each knot propagates.
An irregular pattern of knots forms with detectable proper motion in a few years and with the spatial separation between consecutive knots being neither uniform in space nor constant in time. 
Differences in conditions in both lobes leads to differences in how they interact and thus differences between the two lobes. Jet asymmetries are discussed further in Section 5.2.

\begin{figure}[h!]
   \includegraphics[width=14cm, angle=-90]{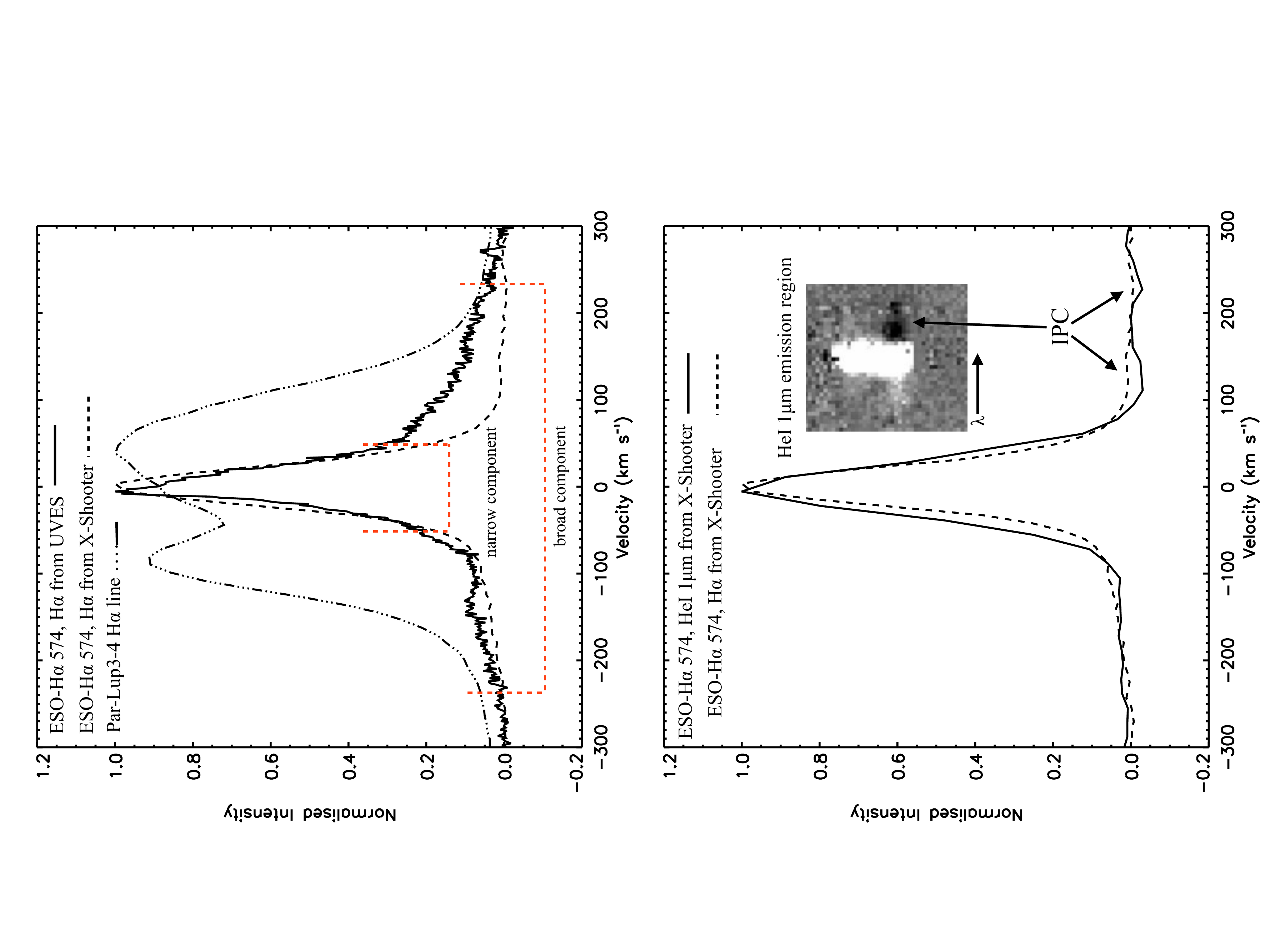}
     \caption{Top Panel: \eso \Ha\ line profiles as observed with \xsh and UVES compared to the \para \Ha\ line profile. The broad and narrow components of the \eso line are marked.  The red-shifted wing of the \eso profile is much enhanced in the UVES data. Bottom Panel: \eso line profiles of He I at 1.083~$\mu$m and \Ha\ (as observed with \xsh) compared to illustrate the presence of an IPC in the He I emission region. This is also seen in the 2D spectrum of \eso (inset). In the section of the 2D spectrum shown, the continuum has been removed and as well as the IPC region and the jet region, faint blue-shifted emission is observed to the left of the jet.}
 \label{Hacomp}        
\end{figure}

\begin{figure}[h!]
   \includegraphics[width=12cm, angle=-90]{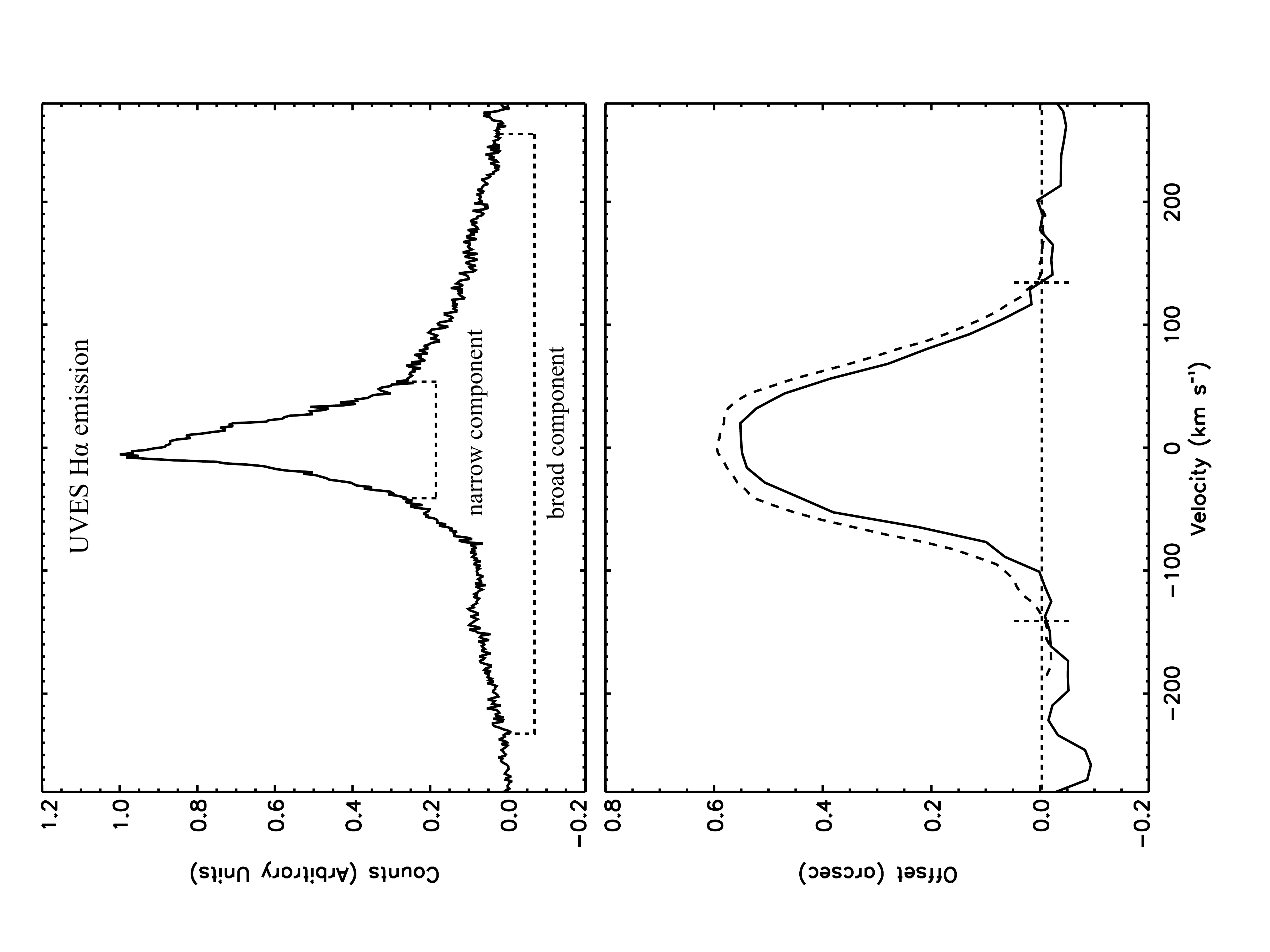}
     \caption{Spectro-astrometry of the \eso \Ha\ emission in the UVES spectrum. Upper Panel: UVES H$\alpha$ line profile is shown and the narrow and broad components of the line are marked. Lower Panel: Spectro-astrometric results are shown and the solid line is the analysis before continuum subtraction and the dashed line after continuum subtraction. The 1-$\sigma$ error is $\sim$20~mas. The H$\alpha$ line is offset in the outflow between $\pm$ 140~\km\ and at least some of the broad component is found to be coincident with the stellar position.}
 \label{SA}        
\end{figure}

\subsubsection{\Ha\ line profile}
As reported in \cite{Bacciotti11}, the \Ha\ emission line region is made up of a narrow and broad component with the former made up of emission from the jet. These components are also detected in the UVES data (see Figure \ref{Hacomp}) and in the He I line at 1.083~$\mu$m (Figures 2, 5). Here the origin of the broad component is considered. Comparing the \Ha\ red-shifted wing in the two observations of \eso points to some red-shifted absorption in the \Ha\ line in the \xsh observation.  The  HeI~1.083~$\mu$m line also exhibits such a red-shifted absorption or inverse P Cygni profile (IPC, see Figure \ref{Hacomp}). This is a signature of infall onto the star \citep{Fischer08}. Studies have shown that the profiles of permitted lines such as \Ha, He I 1.083~$\mu$m, and Pa$\beta$ can exhibit a variety of shapes which are explained by models incorporating both accretion and ejection scenarios \citep{Whelan04, Kurosawa06, Fischer08}.  While definite signatures of accretion and ejection such as IPCs and P Cygni (PC) profiles are seen in permitted lines some of the time, the broad width of the lines and high velocity line wings are a constant feature and high velocity symmetric wings up to $\sim$ 1000~\km\ have been detected in the \Ha\ lines of TTSs  \citep{Reipurth96}. Spectro-astrometry has been used to disentangle accretion and outflow components to permitted emission lines and has shown that the wings of the \Ha\ line emission can originate in an outflow or wind \citep{Takami01, Whelan04}.

The case of \eso is complex as the normally strong accretion tracers are very significantly suppressed and the emission line regions are dominated by the emission from the narrow jet. This is the opposite of what is normally seen in YSOs. Therefore it is difficult to say if the broad component is part of the accretion contribution to the total \Ha\ emission region that becomes visible at the edges where the jet is no longer dominant, or is in fact a true line-wing tracing a wind.  In Figure \ref{Hacomp} (top panel) the \Ha\ line profiles of \eso are compared to that of \para. As the shape of the \para \Ha\ line profile is consistent with an origin in the accretion flow \citep{Bacciotti11}, this comparison is done to illustrate the more complex nature of the \eso line profile. Note that the width of the \eso broad component is comparable to the width of the \para H$ \alpha$ line suggesting that the broad component is perhaps more consistent with an accretion flow.

\cite{Bacciotti11} performed a spectro-astrometric analysis of the wings of the ESO-H$\alpha$ 574 and Par-Lup 3-4 H$\alpha$ lines to check for an outflow contribution. Spectro-astrometry is a technique by which Gaussian fitting is applied to the point spread function (PSF) of a spectrum to measure spatial shifts with respect to the centroid of the continuum position. The accuracy to which these offsets can be measured depends on the signal-to-noise ratio \citep{Whelan08}. The wing emission was found to be coincident with the continuum position and the analysis showed that any offset would be below the error in the analysis of $\sim$ $\pm$ 40~mas for ESO-H$\alpha$ 574 and $\pm$ 20~mas for Par-Lup3-4.  In Figure \ref{SA} the spectro-astrometric analysis of \cite{Bacciotti11} is performed for the \Ha\ emission region in the UVES spectrum. The full line is the offset before continuum subtraction, the dashed line after continuum subtraction and the 1-$\sigma$ error in the centroid measurements is $\sim$ 20~mas. The region between +140~\km\ and -140~\km\ is offset from the continuum position and therefore this region is dominated by emission from the outflow. The rest of the broad component beyond 140~\km\ is coincident with the continuum position and therefore is likely dominated by emission from the infall. Overall we conclude that the broad component of the \eso \Ha\ region is likely made up of emission from infall and outflow. 
Also the comparison between the \eso \xsh and UVES data does point to variability in the \Ha\ emission region. This is based on the fact that the red-shifted wing is not detected in the \xsh spectrum. Variable accretion and outflow activity is a typical characteristic of YSOs \citep{Costigan12}.


 \begin{figure}[h!]
   \includegraphics[width=9cm, trim= 0.2cm 1cm 0.5cm 1cm, clip=true]{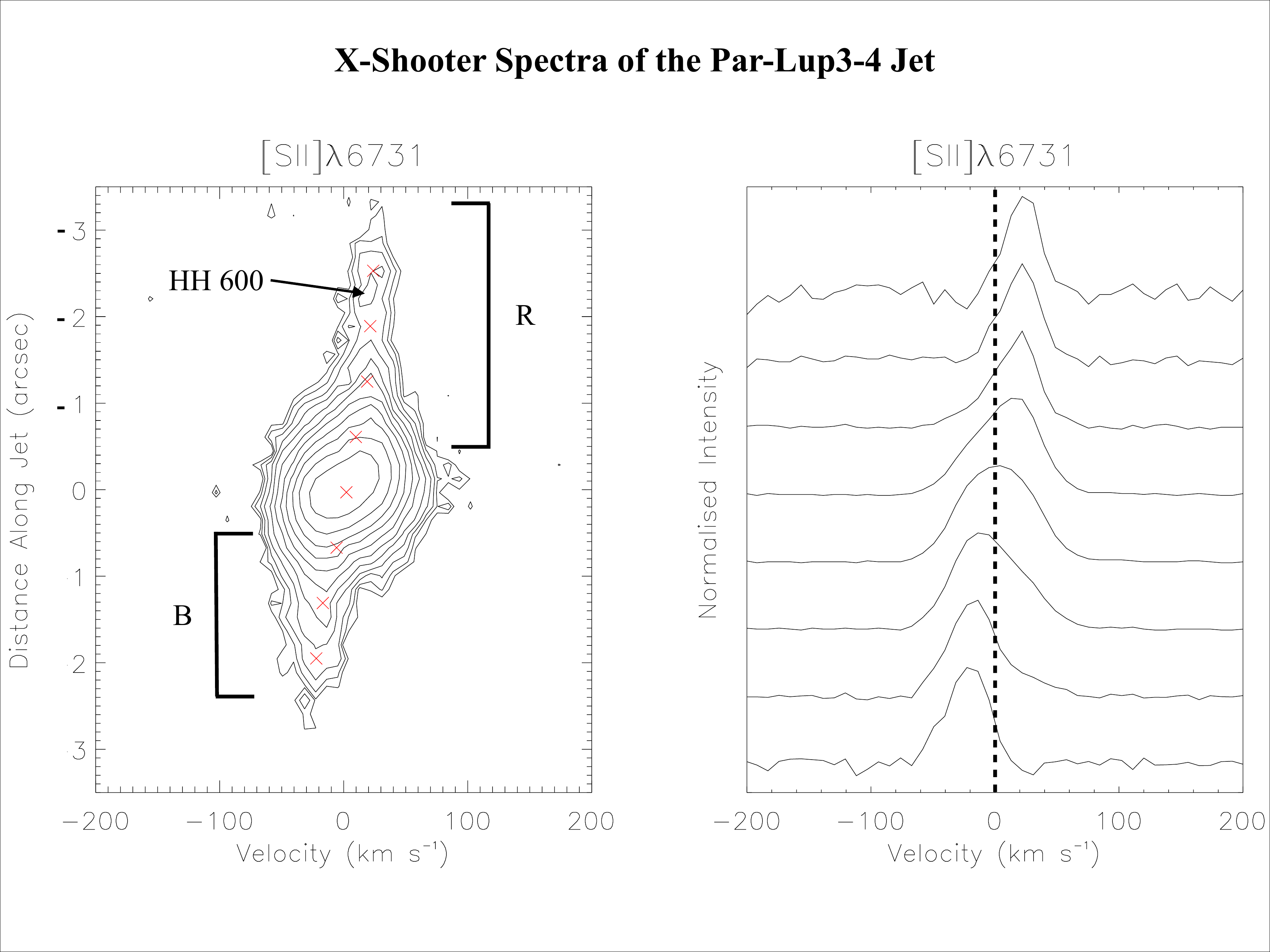}
     \caption{Position velocity diagram of the \para jet in [S II]$\lambda$6731 with the velocity centroid measured at various positions along the jet marked (red crosses). Right panel: line profiles corresponding to each red cross. Contours begin at 1.5 $\times$ 10$^{-18}$ ergs/s/cm$^{2}$ and increase logarithmically to 68 $\times$ 10$^{-18}$ ergs/s/cm$^{2}$. Velocities are systemic. The extraction window for the spectra of the blue and red jets are also shown. The fluxes of the lines which are detected in these regions of the jets are given in Table A2. The \para jet is fairly symmetric and clearly the velocity of both lobes increases with distance from the source. \cite{Comeron11} note the presence of a small red-shifted knot at $\sim$ -2\farcs5 which is also seen here.}
 \label{parajet}        
\end{figure}

\subsection{Kinematics of the \para Jet} 

The low radial velocities measured in the \para jet ($\pm$ 20~\km) are consistent with the small inclination angle of the jet. 
 Figure \ref{parajet} presents the PV diagram of the \para jet in the [SII]$\lambda$6731 line with the red crosses marking velocity centroid measured at fixed intervals along the jet. The line profiles corresponding to each red cross are also shown. The jet which extends to $\pm$ 3~\arcsec\ on either side of the source displays no asymmetries in agreement with \cite{Comeron11}. \cite{Comeron11} compare images of this jet with earlier observations and point out that the appearance of the jet is much more symmetric than in previous observations with both the blue and red-shifted lobes displaying similar intensities. They report the fading of the knot HH~600 between their 2003 and 2010 observations. HH~600 was observed in 2003 in the red-shifted lobe at a distance of 1\farcs3 from \para \citep{Comeron05}. HH~600 is seen here in Figure \ref{parajet} at a distance of $\sim$ 2\farcs5. 
 It is clear from Figure \ref{parajet} that the radial velocity of the jet increases with distance from \para. An increase in radial velocity with distance from source has commonly been observed for jets driven by CTTSs \citep{Davis03} and is a feature of a magnetically driven flow \citep{Whelan04}.

\subsection{Estimates of extinction}

An accurate estimate of the extinction is necessary for measuring $\dot{M}_{acc}$ (from the luminosity of accretion indicators), and thus the ratio $\dot{M}_{out}$ / $\dot{M}_{acc}$. However, for sources possessing an edge-on disk, like the ones under investigation here, this can be complicated by the scattering from the disk material, which can produce  {\it grey extinction}, 
i.e. light suppression independent of wavelength \citep{Mohanty07}. For ESO-H$\alpha$~574, \cite{Luhman07} estimated A$_{\rm J}$=0.45\,mag
meaning that the visual extinction (A$_{\rm V}) \approx$1.7\,mag.
Literature values of A$_{\rm V}$ for Par-Lup3-4
range from 2\,mag to 5\,mag, depending on the colours
used to estimate reddening and an average of $\sim$3.5\,mag has been used in other studies \citep{Comeron03}. In \cite{Bacciotti11}, using the above results, we adopted 
A$_{\rm V}$=3.5\,mag and A$_{\rm V}$=1.7\,mag for Par-Lup3-4 and ESO-H$\alpha$~574, 
respectively.

One goal of the present work was to recover more accurate values for the 
extinction of both sources. This was done by comparing their spectra 
with the X-Shooter spectra of young zero-extinction 
Class III templates \citep{Manara13} of the same spectral type. 
The match was performed by applying to the template (i) a constant factor to normalise
the flux  and (ii) an reddening law to impose an artificial reddening and reproduce the slope of the observed spectrum. 
If the latter correction is zero, then extinction is most likely grey. 
Uncertainties in spectral types and in extinction of the templates lead
globally to an error of $<$0.50\,mag. Note however that veiling 
(especially strong in \eso) and dust 
scattering both increasing toward the blue, may still lead
to an underestimation of the extinction. See also \cite{alcala13} .

The comparison was carried out as follows. The templates were normalised 
to the flux at 750\,nm and then artificially reddened between
A$_{\rm V}$ = 0...4.0\,mag, in steps of 0.25\,mag, until the best match
to the Par-Lup3-4 and ESO-H$\alpha$~574 spectra was determined. 
It was found that a relative extinction correction  
of 1.5\,mag is needed to best match the K8-type template to the spectrum 
of ESO-H$\alpha$~574, whereas no relative extinction is necessary 
to match the M5-type template to the spectrum of Par-Lup3-4, 
suggesting that \para is dominated by grey extinction.
The extinction-corrected luminosity 
of ESO-H$\alpha$~574, however, is still considerably lower than the average luminosity 
of other YSOs of similar mass in Cha~I-north.  We 
argue therefore that \eso is also affected by grey extinction. This is supported by the similarity of this system with the one of TWA30 described in  
\cite{Looper10a, Looper10b}. There is evidence that TWA 30A has an edge-on disk and it exhibits similar spectra to ESO-H$\alpha$ 574, in that strong outflow tracers are detected but accretion tracers are relatively weaker.

In principle the adopted normalisation factor should be a measure of the 
obscuration of the source by the intervening disk. This could in fact be obtained 
by multiplying the flux ratio of the object to the template 
by the square of the ratio of their distances, assuming a similar stellar radius.
The template and the object, however, are generally not in the same 
evolutionary stage and hence have different radii. 
Also, the uncertainties in distances are large. 
Therefore we keep with the approach of \cite{Bacciotti11} and 
adopt a `grey' obscuration factor of approximately 150 and 25 for 
ESO-H$\alpha$~574 and Par-Lup3-4, respectively. \cite{Bacciotti11} obtained these values by comparing 
luminosities corrected for any relative extinction 
to the average luminosities of young 
stars of similar mass in the same star forming regions. 
Regarding the forbidden lines generated in the jets, however, 
we note that  from analyses of infrared lines 
of Fe$^+$, \cite{teresa13} find that these regions are not 
affected by extinction. Therefore A$_V = 0$ is assumed for
the diagnostics of physical conditions in the jets of both sources.



\subsection{Estimates of $\dot{M}_{acc}$}

As discussed in the Introduction, previous estimates of $\dot{M}_{acc}$ for both \eso and Par-Lup3-4 were re-analysed.
$\dot{M}_{acc}$ is calculated from the equation, 
\begin{equation}
\dot{M}_{acc} = 1.25 (L_{acc} R_{*}) / (G M_{*})
 \label{Macc}
\end{equation}
\noindent 
where L$_{acc}$ is computed from L$_{line}$ \citep{Gullbring98}.  L$_{line}$ is the line luminosity derived from the measured fluxes after correction for relative extinction, and $R_{*}$, $M_{*}$ are the stellar radius and mass. For both
sources L$_{line}$ was estimated from the line fluxes reported in
Tables A1 and A2. So for \eso it is the luminosity of the lines
extracted from the region of knot A1. $M_{*}$  and $R_{*}$ are taken at 0.5~\Msun\ and
0.13~\Msun, and at 0.12~\Rsun\ and 0.18~\Rsun, for \eso and Par-Lup3-4,
respectively.
The  values of L and T$_{eff}$ used in the computation of R$_{*}$ are
taken from \cite{Merin08} and \cite{Luhman07}.

The new estimates differ from those presented in 
\cite{Bacciotti11} due to the following factors. Firstly, improvements in the flux calibration mean that line fluxes and therefore values of  L$_{line}$ have changed. Early versions of the \xsh pipeline did not offer a reliable flux calibration but this situation has since been resolved. This was more of an issue for ESO-H$\alpha$ 574, as fluxes are now found to be lower 
by a factor 4-5. Improvements     
in background and skyline subtraction, 1-D extraction, and telluric correction also led to more accurate measurements of L$_{line}$. Secondly, for the calculation of L$_{line}$ only the line emission close to the source, i.e. 
knot A1 in the case of \eso and the arcsecond centred on the star in the case of Par-Lup3-4, is considered. Previously, the emission extracted over the whole spatial range of the spectrum was considered. Thirdly, the new estimates of relative extinction described in the above paragraph have been applied using the extinction curves of \cite{Weingartner01}. Finally, for the calculation of L$_{acc}$ from L$_{line}$ the new relationships described in \cite{alcala13} are adopted. These relationships
consider a combination of all the accretion indicators calibrated on sources for which 
the Balmer jump has been measured simultaneously. 

\begin{figure}[h!]
   \includegraphics[width=9.5cm, trim= 1.5cm 1.2cm 0cm 0.0cm, clip=true]{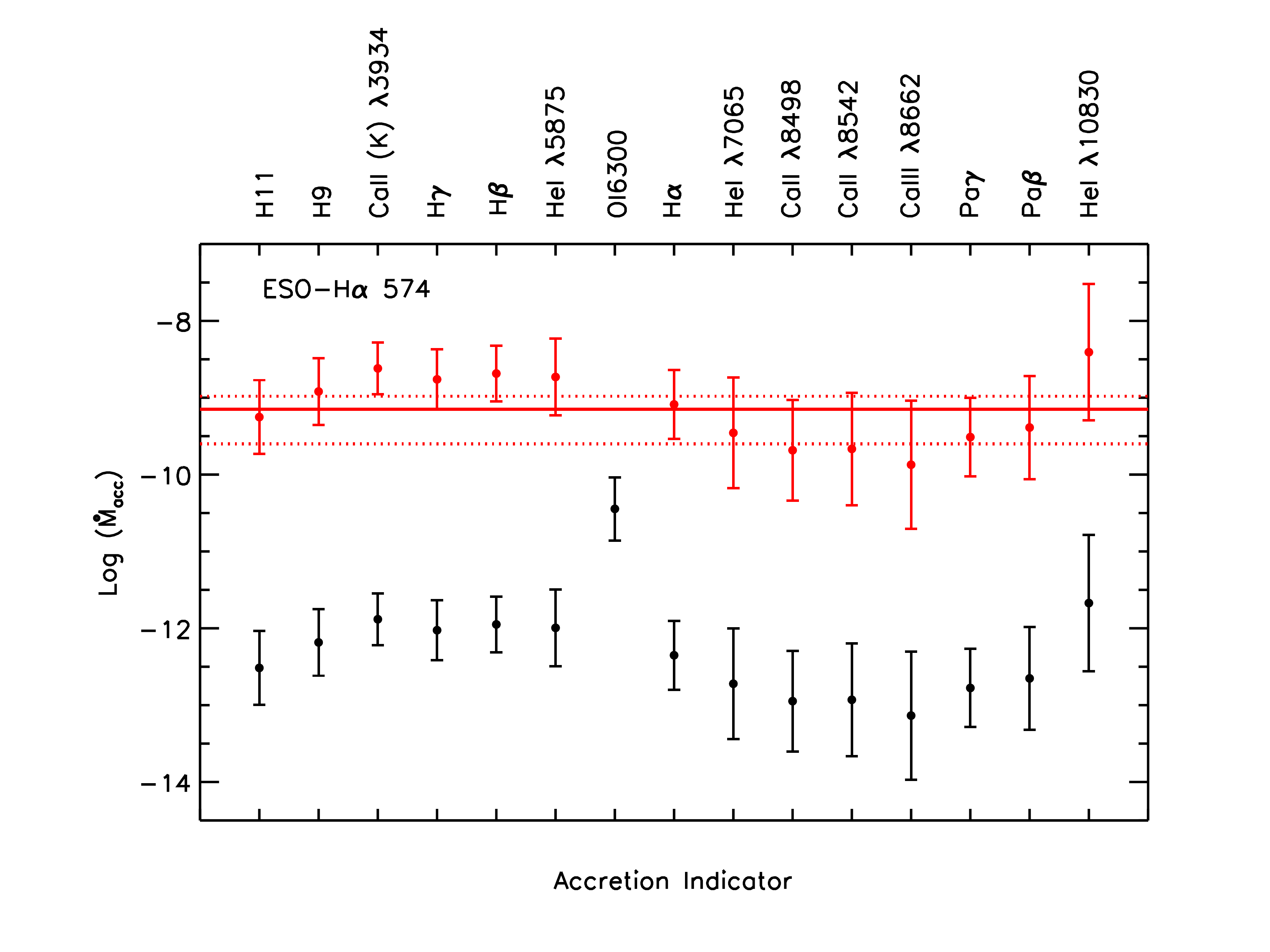}
    \includegraphics[width=9.5cm, trim= 1.5cm 1.2cm 0cm 0.0cm, clip=true]{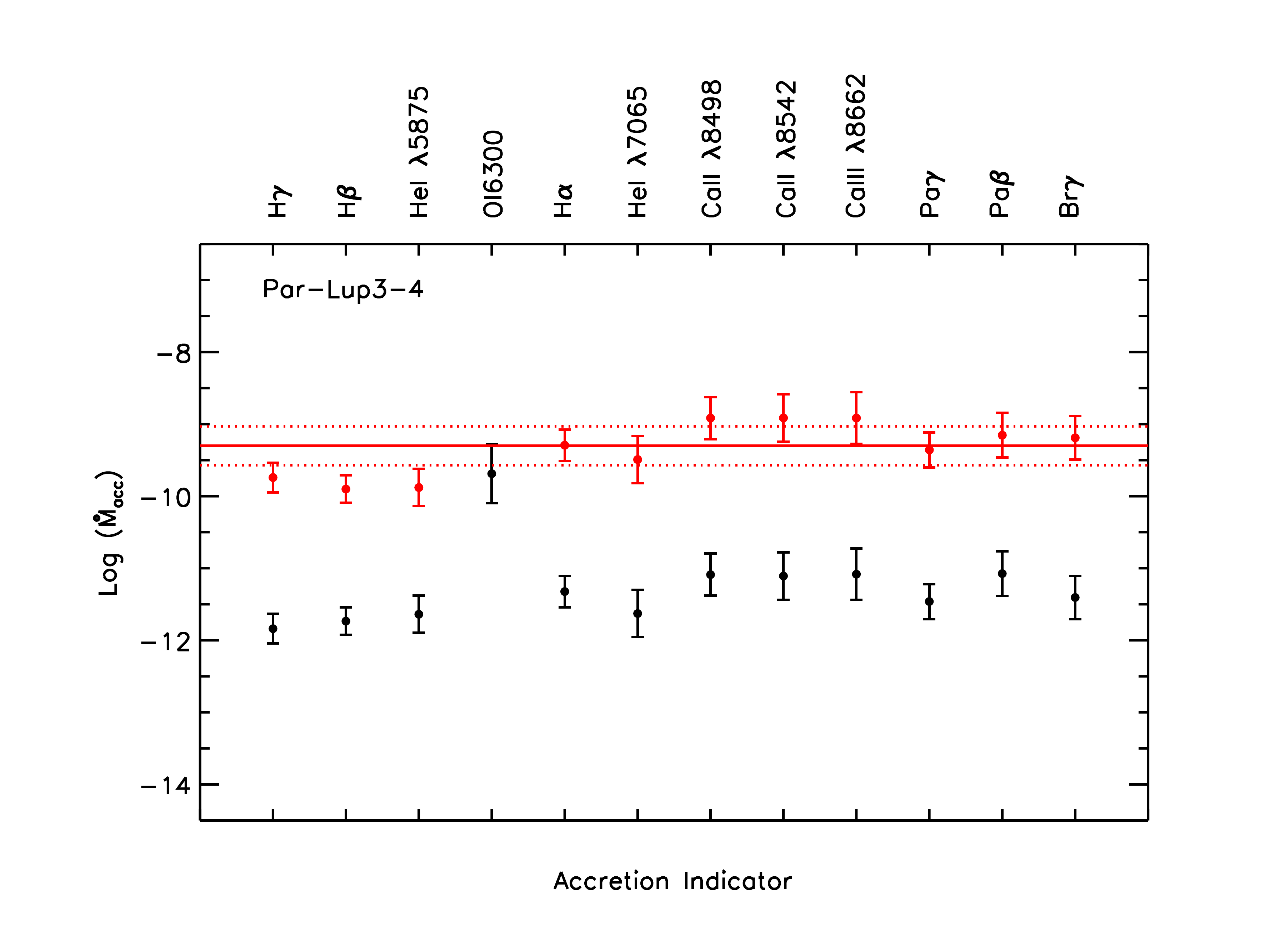}
     \caption{Mass accretion rate calculated for various accretion indicators in \eso (top) and Par-Lup3-4 (bottom). The red points are the values of $\dot{M}_{acc}$ after correction for obscuration by the disk. [OI]$\lambda$6300 emission is an indirect indicator of accretion coming from extended regions and as such is blocked by disk. Therefore it is used as a test of the factor chosen to correct the other tracers for disk  obscuration. The solid line gives the mean value of log($\dot{M}_{acc}$) and the dashed line is the $\pm$ 1~$\sigma$ uncertainty.}
 \label{accretion}        
\end{figure}

The obscuration effect of the
edge-on disks of both sources causing grey extinction 
is corrected for in the following way.
At fixed mass Equation~\ref{Macc} implies $\dot{M}_{acc}$~$\propto L_{acc}\cdot L_{*}^{0.5}$,
as the stellar radius goes as the square root of the luminosity. 
Assuming that the same obscuration factor suppresses both  $L_{acc}$ 
and $L_{*}$, $\dot{M}_{acc}$ is corrected using:
\begin{equation}
\dot{M}_{acc}(corrected) = (obscuration factor)^{1.5}*\dot{M}_{acc}
\end{equation}
\noindent 
where the obscuration factor was taken to be 150 and 25 for \eso 
and \para, respectively, as described in Section 4.4. In Figure \ref{accretion}, $\dot{M}_{acc}$ for \eso and \para is shown, with 
the black and red points resulting from the calculation before and after correction for the obscuration factor. 
Considering the corrected values, results put the mean accretion rate at 
log($\dot{M}_{acc}$)= -9.15 $\pm$ 0.45 \,\Msun~yr$^{-1}$  for \eso 
and  -9.30 $\pm$ 0.27 \,\Msun~yr$^{-1}$  for \para. The errors are the 1-$\sigma$ value marked in Figure 8.
$\dot{M}_{acc}$ for \eso is at the lower end of the range of values of $\dot{M}_{acc}$ 
measured for other K type YSOs  \citep{Gullbring98}, and a factor ten lower than the approximate estimate in 
\cite{Bacciotti11} with obscuration considered. The result for \para is closer to our previous estimate in  \cite{Bacciotti11},
and it is in agreement with previous results for M type YSOs \citep{AntonS11, Herczeg08, Hartigan95}.

These values of $\dot{M}_{acc}$ are compared in Section 4.6.3 with new estimates of $\dot{M}_{out}$ . 
First note however, that additional uncertainties affect  
the $\dot{M}_{acc}$ estimate of ESO-H$\alpha$ 574. For \eso the
accretion indicators are clearly dominated by jet emission. This can
be seen in the PV plots of Figure 2 where the emission peaks are always offset 
and also in the spectro-astrometric analysis presented in Figure \ref{SA}. Also refer to the
discussion of the Balmer decrements in Section 4.6.1 below. This can lead to inconsistencies
because, on one hand, the obscuration factor is estimated from the continuum radiation 
and not from the line emission, and on the other hand, 
the  empirical relationships
between L$_{line}$ and L$_{acc}$, are derived from a sample of 
sources in Lupus \citep{alcala13} where L$_{line}$ is actually dominated by accretion. 
As an attempt to quantify these caveats, $L_{acc}$ is also derived
using a linear correlation with an indirect tracer, i.e. 
the [OI]$\lambda$6300 line \citep{Herczeg08}. The [OI]$\lambda$6300 line is normally a strong tracer of protostellar jets and here we use the fact that empirical correlations between $\dot{M}_{acc}$ and [OI] $\lambda$6300 have shown that [OI] $\lambda$6300 indirectly traces accretion. As the [OI]$\lambda$6300 emission originates mostly in the extended regions, it is not affected by the obscuration of the disk and therefore one could expect $\dot{M}_{acc}$ derived from the [OI]$\lambda$6300 line to agree with the obscuration corrected values of $\dot{M}_{acc}$ derived from the accretion tracers. Thus $\dot{M}_{acc}$ calculated from the [OI]$\lambda$6300 line is a test of the accuracy of the obscuration correction.
However,  
it should be kept in mind that the [OI]$\lambda$6300 correlation with L$_{acc}$ is not perfect and \cite{Herczeg08} point out that the extended 
[OI]$\lambda$6300 emission may enter the slit only partially, which could account 
for the correlation scatter.
In Figure~\ref{accretion} it can be seen that the  M$_{acc}$([$\ion{O}{i}$])  
result for \para is perfectly consistent with the corresponding values
derived from the direct accretion tracers, corrected for obscuration.
The comparison is not as good  for ESO-H$\alpha$ 574, however. The discrepancy may be alleviated considering 
that an imperfect sky line subtraction may have  suppressed part of the [OI] flux.  
Furthermore, veiling and scattering close to the source may have actually affected the line, 
while here it is assumed to be seen under zero extinction conditions.


\vspace{0.5cm}

\subsection{Physical properties of the gas and mass outflow rates}

\subsubsection{Diagnostics from hydrogen lines}

\begin{figure}[h!]
\centering
   \includegraphics[width=9.5cm, trim= 3.0cm 0.5cm 0cm 2cm, clip=true]{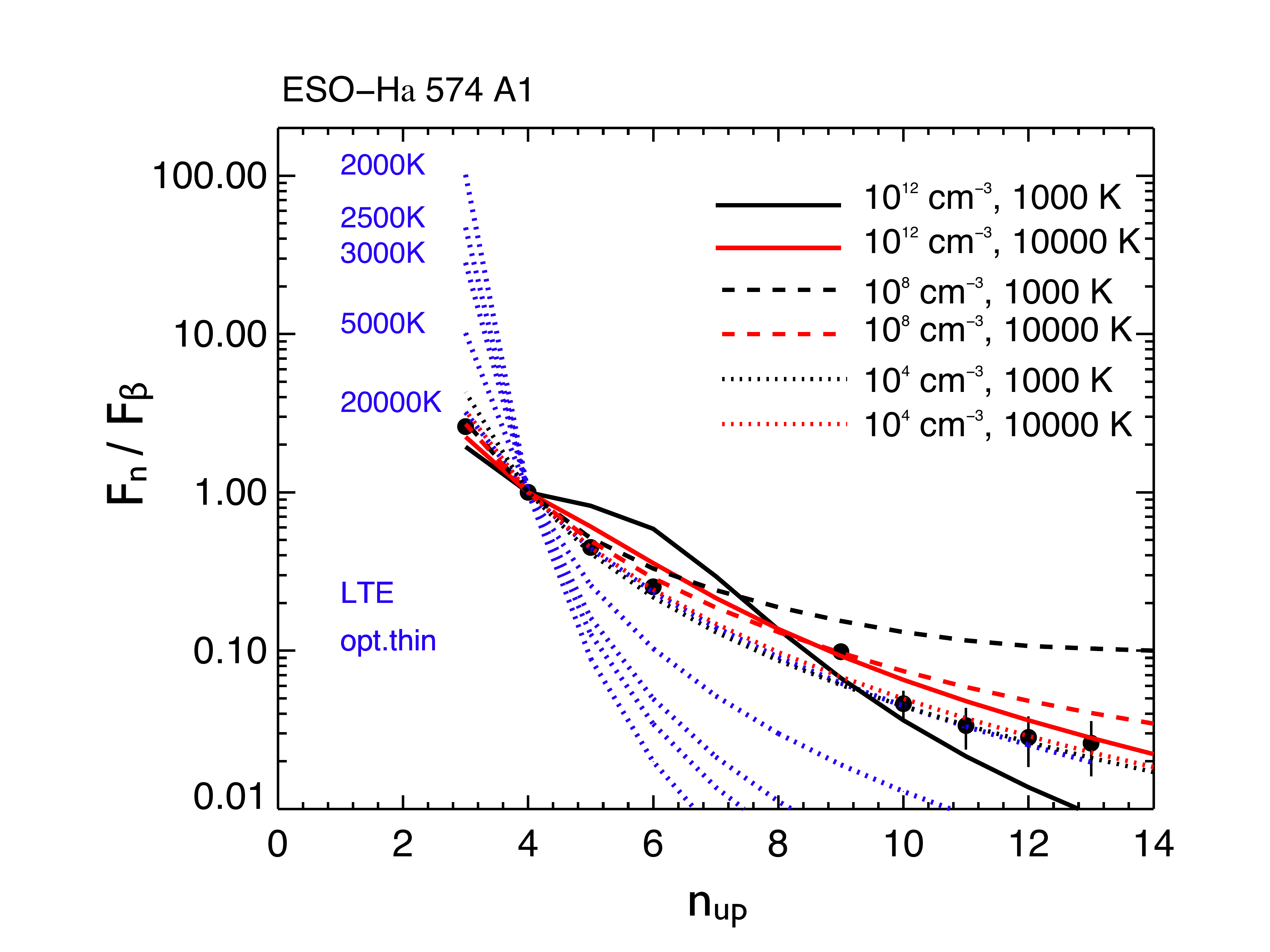}
      \includegraphics[width=9.5cm, trim= 3.0cm 0.5cm 0cm 1cm, clip=true]{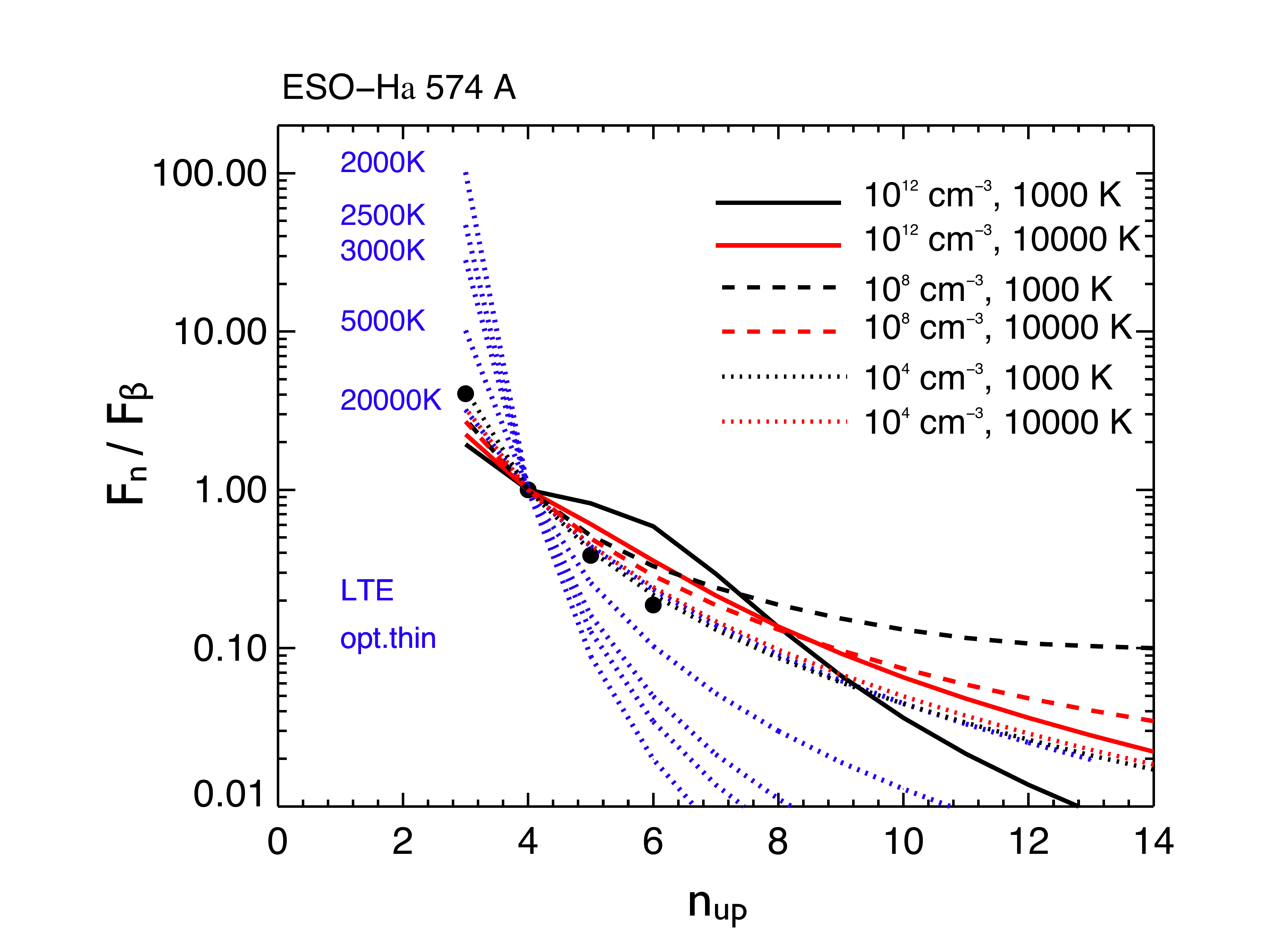}
     \caption{Balmer decrements for ESO-H$\alpha$ 574, regions A1 (top) and A (bottom), fitted with Case B recombination (black and red lines) and LTE optically thin models (blue line). For the Case B scenario, the best fit is offered by n$_{e}$ =10$^{4}$~cm$^{-3}$, T$_{e}$ = 10,000~K. The LTE optically thin case with T$_{e}$ = 20,000~K provides an equally good fit. The results agree with the fact that for region A1, emission from the accretion zone is dominated by the outflow emission. }
 \label{esobal}        
\end{figure}

\begin{figure}[h!]
   \includegraphics[width=9.8cm, trim= 3.0cm 0.5cm 0cm 1cm, clip=true]{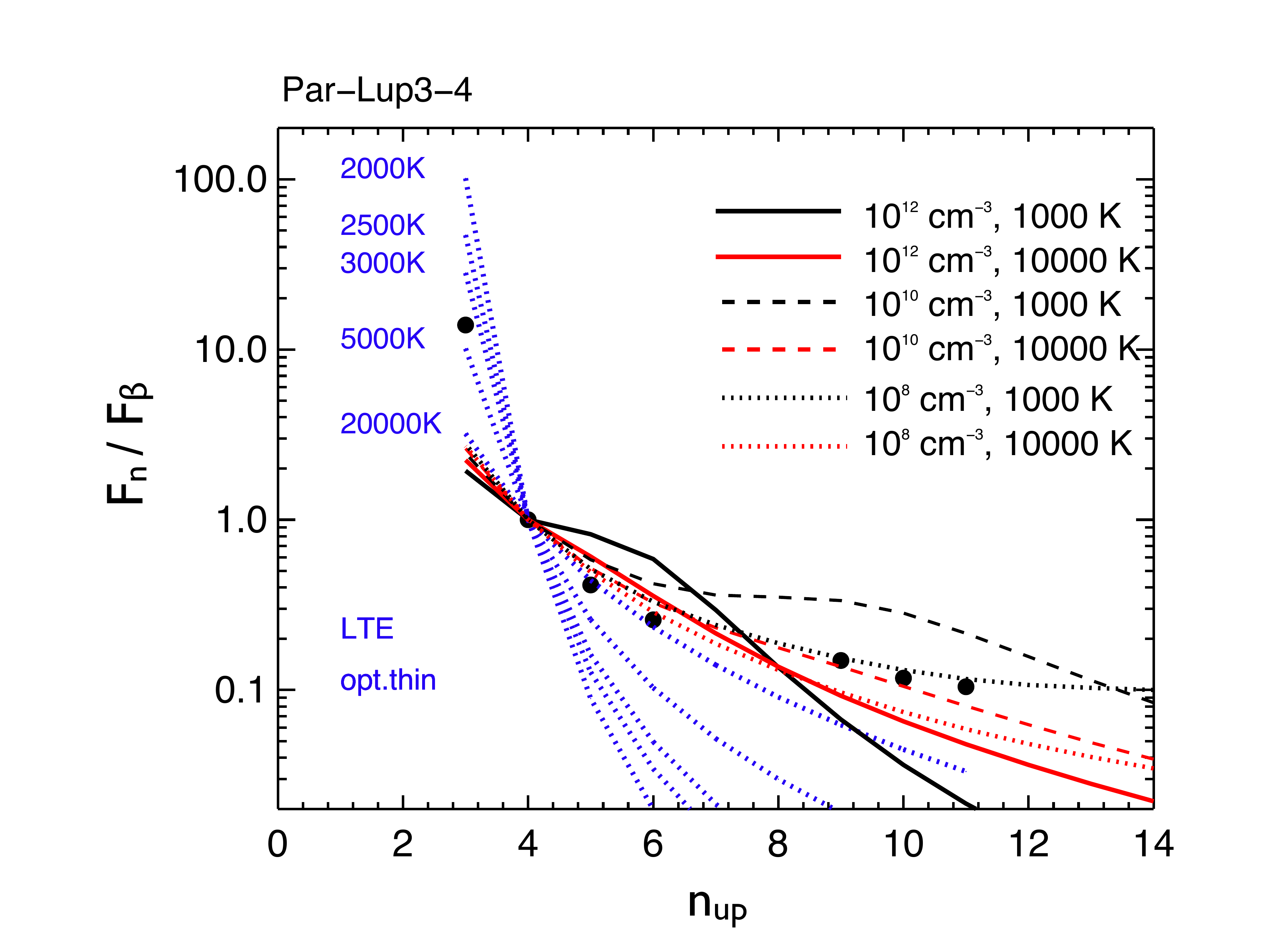}
          \caption{Balmer decrements for \para. Fluxes are extracted from the source spectrum. The decrements are fitted with Case B recombination (black and red lines) and LTE optically thin models (blue lines). The best fit is offered by the Case B scenario and n$_{e}$ =10$^{10}$ cm$^{-3}$, Te = 10,000 K. This is consistent with magneto-spheric accretion.}
 \label{pasbal}        
\end{figure}

In order to understand the physical conditions in the various emitting regions of \eso and \para the hydrogen emission from A1 and A in the \eso flow and from Par-Lup 3-4 was 
investigated. A1 which covers the region -0\farcs5 to 1\farcs5 also includes the emission from \eso itself. Numerous HI emission lines are detected in the spectra of both sources and the Balmer decrements for \eso A1, A and \para were computed (with respect to H$\beta$). The line ratios are plotted as a function of their upper quantum number (n$_{up}$) and results are presented in Figures \ref{esobal} and \ref{pasbal}. 
While the spectral range of \xsh covers all of the Balmer lines, any lines with n$_{up}$ $>$ 13 were found to be too noisy for inclusion in the analysis. Additionally, at the intermediate spectral resolution of \xsh the H7 and H8 lines are blended with other lines and are therefore also not included. It is noted that only the brightest Balmer lines were detected in knot A.
To constrain the physical conditions in the emitting gas the decrements were firstly compared to standard Case B predictions (red and black curves) calculated for a range of temperature and density. The Case B curves were derived using the calculations of \cite{Hummer87} and the data files provided by \cite{Storey95}. These models assume that all lines are optically thin. Secondly, the decrements were also compared to optically thick and thin local thermodynamic equilibrium cases (LTE, blue curves) ratios, calculated over a temperature range of 2000~K to 20000~K. The optically thick case turned out not to be applicable, 
as it did not provide a good fit to the results, and it is not shown here.

For regions A1 and A in the \eso outflow the Case B model for n$_{e}$ = 10$^{4}$~cm$^{-3}$, T$_{e}$ = 10000~K is a 
good fit to the decrements, as well as, for A1, the LTE optically thin curve with T$_{e}$ = 20000~K. 
A density of $\sim$ 10$^{4}$~cm$^{-3}$ and a temperature range of 10,000-20,000~K is consistent with an origin in 
a jet and also agrees with results published in other jet studies  \citep{Podio10, Podio08}. 
This analysis shows that the bulk of the Balmer emission from A1 comes from the outflow, in agreement with the
appearance of the PV diagrams of Figure \ref{mix} and the spectro-astrometric analysis of the \Ha\ line for this source (Figure 6).  

The gas from \para is constrained with the Case B models, a high density of 10$^{8}$-10$^{10}$ cm$^{-3}$ and a temperature of $\sim$ 10,000~K. Such temperatures and densities agree with an origin in magnetospheric accretion columns \citep{Martin96}. For example, \cite{Muzerolle01} investigated magnetospheric accretion in TTS and limited the temperature range of magneto-spherically accreting gas at 6000~K $<$ T $<$ 12000~K , for 
10$^{-6}$ \Msun yr$^{-1}$ $\geq$ $\dot{M}_{acc}$  $\geq$ 10$^{-10}$ \Msun yr$^{-1}$, where the 
lower gas temperatures correspond to sources with higher mass accretion rates. 
The origin of the \para HI emission in accreting gas is in agreement with 
the spectro-astrometric analysis of the \Ha\ line \citep{Bacciotti11}.

The temperature estimated for \para disagrees with the results of \cite{Bary08} who found that the temperature of the gas emitting the Paschen and Brackett lines, in a sample of TTSs, was best fitted with a value of $\lesssim$ 2000~K. They argue that while this value is well below the temperature predicted by magnetospheric accretion models, the models do not currently allow for the absorption by the gas of high energy photons from the hot corona and/or accretion shocks. Inclusion of this method of heating and ionisation means that the low gas temperature can still be consistent with HI emission arising from accreting gas, as the ionising photons allow for the production of intense HI emission even at low temperatures. The majority of sources in the sample of \cite{Bary08} have values of $\dot{M}_{acc}$ on the order of 10$^{-7}$ \Msun yr$^{-1}$, corresponding to the lower end of the temperature scale proposed by \cite{Muzerolle01}. 
While \para has a lower value of $\dot{M}_{acc}$, which suggests higher temperatures
than for the sources in the study of \cite{Bary08}. 

\subsubsection{Diagnostics From ionic lines}

\begin{figure*}
\centering
   \includegraphics[width=16cm, trim= 0.0cm 2cm  0.0cm 2cm, clip=true]{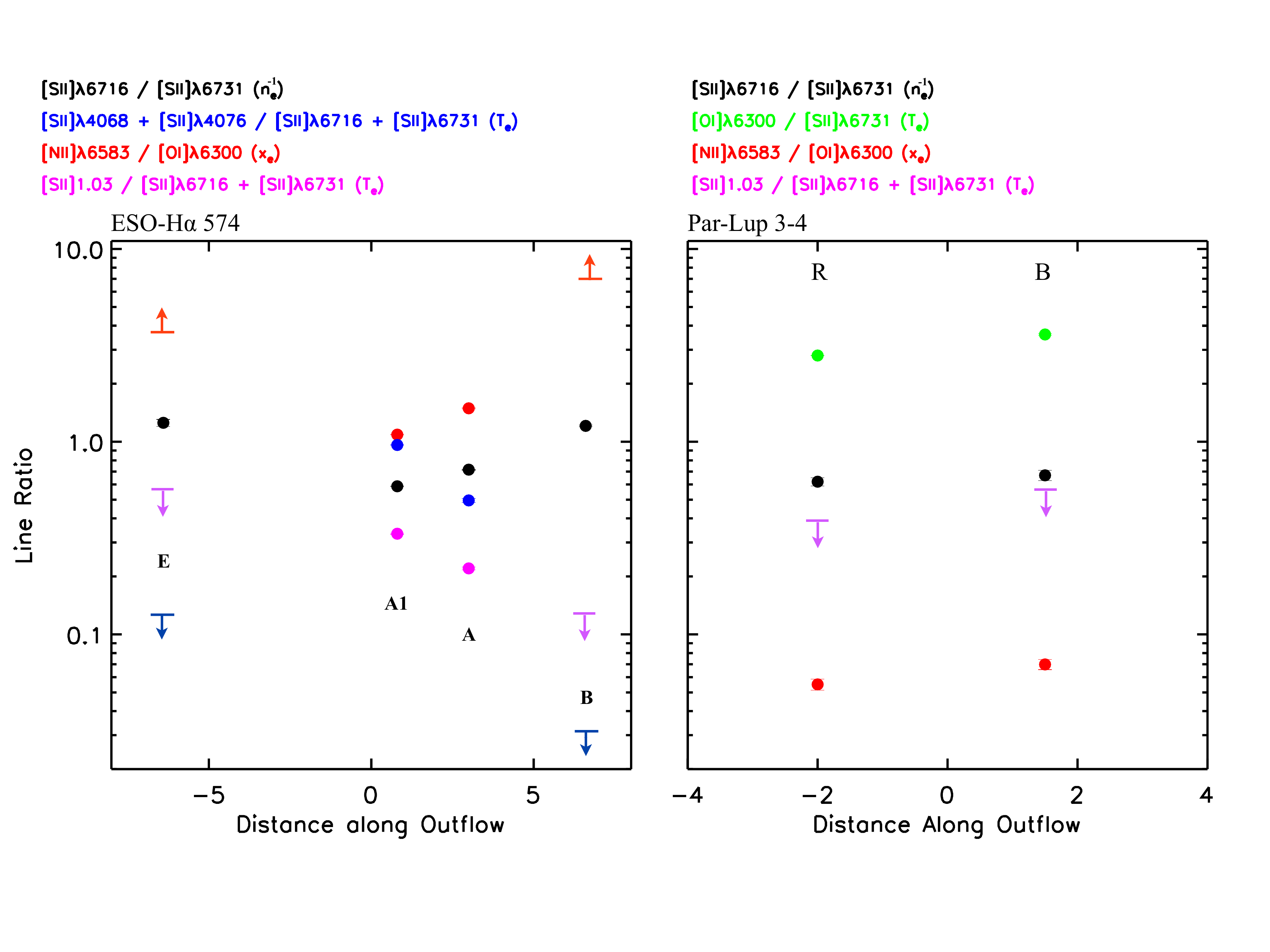}
     \caption{Various line ratios tracing the excitation, n$_{e}$, T$_{e}$ and x$_{e}$ of the gas in \eso A1, A, B and E and the \para blue and red jets are plotted. 
As some lines were not detected in B and E and in the \para jets upper limits of 3-$\sigma$ were assumed.} 
\label{ratio}           
\end{figure*}

In Giannini et al. (2013) the numerous [Fe II] and [Fe III] lines detected in our objects 
were used to derive the physical parameters (n$_e$, x$_e$ and T$_e$) at the base of the \eso and \para jets.
This paper pointed out the large gradients of physical conditions present in the post-shocked gas and
that different lines/species probe different gas layers. Here, to further probe the gas in the two jets 
we employ the other line ratios 
sensitive to the gas physical conditions. 
The gas physical conditions are determined through a comparison of the observed values of the line ratios with values 
predicted theoretically using a numerical code that calculates the level population in a 5-level model atom \citep{BE99, Podio06}. Not all the lines, however, are present in a given knot, thus 
different combinations are used in the various positions. 
The values of the considered ratios or their upper/lower limits are 
plotted in Figure \ref{ratio} for knots A1, A, B and E of \eso and for the jets in Par-Lup3-4, and the results are given in Table 1.

For the \eso outflow 
a combination of [SII] lines were used to estimate $n_e$ and $T_e$ independently from the abundance  
and ionisation of S. In particular for A1 and A
the [SII]$\lambda$6716 /  [SII]$\lambda$ 6731 ratio, dependent strongly on $n_e$,   
and the ([SII]$\lambda$4068 +[SII]$\lambda$4076) / ([SII]$\lambda$6716 + [SII]$\lambda$6731) 
ratio, dependent strongly on $T_e$, were used. Values  of $T_e$ around $1.4~10^4$ K, and $n_e \sim$ 4500, 2100  cm$^{-3}$  were measured in A1 and A, respectively (see Table 1).
Assuming the same temperature in the outermost knots 
we obtain $n_e \sim$ 230 cm$^{-3}$ in B and  160 cm$^{-3}$ in E.  
Errors are 15\% on the temperature and 5\% on the electron density.
The ratio [NII]$\lambda$6583 / [OI]$\lambda$6300, proportional to hydrogen ionisation fraction $x_e$,
indicates $x_e > 0.65$ in the flow, and  increases moving away from the source. 
This is in agreement with the value found on-source by Giannini et al. (2013). 
The high excitation of B and E would seem to contradict the non-detection of 
lines from higher ionisation lines such as [SIII] [OIII] in these knots (see Figure 2), 
however we argue that this is due to the low electron density, not high enough 
to collisionally excite these lines. 
Note that the values retrieved on knot A1 are consistent with the estimates given 
in \cite{teresa13} for the analysis of the [SII] lines. 
At variance, iron lines, probing gas at different excitation
in the post-shocked gas, provide a range of values for these parameters.
Temperature (density) values from the [SII] lines are at the upper (lower) end of values derived
from the iron lines. 
%

For the \para case the ratio 
[OI]$\lambda$6300 / [SII]$\lambda$6731 was used in conjunction with  [SII]$\lambda$6716 / [SII]$\lambda$6731 and 
[NII]$\lambda$6583 / [OI]$\lambda$6300, to find $n_e, T_e, x_e$ within the framework of the so-called BE technique 
\citep{BE99, Podio06}. Here an improved BE procedure was used which
considers the variation of $T_e$ with the [SII]$\lambda$6716 / [SII]$\lambda$6731ratio. 
Note that the BE technique is not applicable to the \eso outflow  
because of the presence of higher ionisation 
states of O and S close to the source, 
and absence of [OI]$\lambda$6300  in knots B and E.
The results for the \para jet at the source position and in the blue and red lobes are given in Table 1. 
Again the results are in agreement with the 
estimates of \cite{teresa13} deduced from [SII] lines.

For both sources the T$_{e}$ sensitive ratio
[SII] 1.03 / ([SII]$\lambda$6716 + [SII]$\lambda$6731) was also investigated. While only upper limits are detected for Par-Lup3-4, overall results are consistent with the
temperature being higher for \para than for the \eso outflow. On the other hand, the low level of hydrogen ionisation 
for \para is confirmed by the ratio of the [NI] and [NII] lines. The difference in the x$_{e}$ and T$_{e}$ values between \eso and \para can be explained in the context
of a shock framework,  as described for example in \cite{Hartigan94}. 
With reference to their Figure 1, higher velocity shocks result in
high x$_{e}$ and high T$_{e}$ at the shock front, but also a strong compression
of the gas. As a result of the high density the gas cools
quickly and the temperature of the post-shock region at the angular resolution of
our observations is not very high. Alternatively, in low velocity shocks, where the
gas has a low x$_{e}$, the compression is lower and the cooling is slower, thus
the average temperature in the observed region has a higher valuer.
Following this logic would mean that B and E in \eso are higher velocity 
shocks than  A1, A and the \para jets. 
This explains why \eso has a spectrum much richer in lines from high ionised species, 
such as [OIII] and [SIII], not detected in the \para outflow. 
\cite{Hartigan94} also plot the various line ratios as a 
function of shock velocity (see their Figures 10 to 14). 
We used these models to estimate the shock velocities and 
found that for the \para  jets and \eso inner knots, 
shock speeds of 30~\km\ to 60~\km\ are indicated, while speeds reach $\sim$ 90~\km\ in B and E of \eso. 

\begin{table*}
\begin{center}
\begin{tabular}{cccccccc} 
\hline   
Object &$v_{t}$ (kms$^{-1}$) &$l_{t}$ (\arcsec) &n$_{e}$ (cm$^{-3}$) &T$_{e}$ (10$^{4}$ K) &x$_{e}$   &$\dot{M}_{out}$ (10$^{-10}$ \Msun yr$^{-1}$) &$\dot{M}_{out}$ / $\dot{M}_{acc}$ \\
\hline
ESO HA 574  & & & & &
\\
Knot A1       &130 $\pm$ 30 &2.8     &4500 $\pm$ 225    &1.4 $\pm$ 0.2      &$>$0.65  & 5.0 $\pm$ 1.5  &
\\
Knot A  &130 $\pm$ 30 &2.4          &2100 $\pm$ 105   &"     &"  & 3.7 $\pm$ 1.1  &0.5 (+1.0)(-0.2)
\\
Knot B &220 $\pm$ 50 &2.2 &230  $\pm$ 11.5       &"       &"    &8.2 $\pm$ 2.5  &
\\
Knot E &325 $\pm$ 50 &2.3 &160   $\pm$ 8.0      &"     &"   &3.5 $\pm$ 1.1  &0.3 (+0.6)(-0.1)
 \\
 Par-Lup3-4 & & & & & & &
 \\
 Blue &170 $\pm$ 30 &1.9 &4300 $\pm$ 200 &3.1 $\pm$ 0.5 &(2.0 $\pm$ 0.2)$\times$10$^{-3}$  &0.3 $\pm$ 0.1 &0.05 (+0.10)(-0.02)
 \\
Red &170 $\pm$ 30 &1.8 &4600 $\pm$ 300 &2.4 $\pm$ 0.6  &(6.3 $\pm$ 0.2)$\times$10$^{-3}$ &0.29 $\pm$ 0.09 &0.05 (+0.10)(-0.02)
 \\
\hline                                                              
\end{tabular}
\caption{Values of the different variables used in Equation 3.}  
\end{center}
\end{table*}

\subsubsection{Estimates of $\dot{M}_{out}$ / $\dot{M}_{acc}$}

The values of n$_{e}$ and T$_{e}$ given in Table 1 are now used to estimate $\dot{M}_{out}$. This is done through a comparison of the observed and theoretical line luminosity 
produced in a given forbidden line. To this aim the luminosity $L_{SII}$ of the [SII]$\lambda$6731 line 
was considered in the separate knots making up the \eso outflow and in the \para blue and red jets. Following \cite{Nisini05,Podio06} 
we use the relationship  
\begin{equation}
\dot{M}_{out} = \mu\,m_{H}\,(n_{H}\,V)\,v_{t}/l_{t}
\end{equation}
with 
$n_H\,V = L_{SII}\,\left(h\,\nu\,A_{i}\,f_{i}\,\frac{X^i}{X}\,\frac{X}{H}\right)^{-1}$. Here $\mu=1.24 $ is the mean atomic weight, $m_{H}$ the proton mass,  
$V$ the volume effectively filled by the emitting gas,
$v_{t}$ and $l_{t}$ the tangential velocity and length of the knot, 
$A_{i}$ and $f_{i}$ the radiative rate 
and upper level population relative to the considered transition and finally
$\frac{X^i}{X}$ and $\frac{X}{H}$ are the ionisation fraction and the relative 
abundance of the considered species. The tangential velocities of the knots as estimated in \cite{Bacciotti11} are used here and the knot lengths are given in Section 4.1. These values are also given in Table 1.
The S=S$^+$, elemental abundance is taken from \cite{Asplund05}, and the level population 
is calculated numerically as described in \cite{Podio06}.
Note that the new estimate is more accurate than in \cite{Bacciotti11}, firstly due to the new values of the observed fluxes, that originate from an improved treatment of the 
flux calibration and the new extinction estimates. Secondly, and most importantly $\dot{M}_{out}$ is now calculated using the 
jet parameters derived directly from the spectra, rather than from an approximate formula based on the value of the critical density at an assumed unknown temperature, as was done previously for outflows from stars of very low mass. 

These updated measurements of $\dot{M}_{out}$ and thus $\dot{M}_{out}$ / $\dot{M}_{acc}$ are given in Table 1. $\dot{M}_{out}$ / $\dot{M}_{acc}$ is derived for each lobe of the \eso and \para outflows by dividing by the average value of $\dot{M}_{acc}$. 
For the \eso blue outflow, the average $\dot{M}_{out}$ of A1, A and B is used. 
From Table 1 it can be seen that in the case of \eso the total (two-sided) $\dot{M}_{out}$ / $\dot{M}_{acc}$ is higher than the maximum 
predicated by current models. However, errors are large for this source and may bring back the
ratio within the limits. Furthermore as mentioned in Section 4.5 it is likely that the effects of the edge-on disk of \eso are still not full understood. Thanks to our revised analysis, the total (two-sided) $\dot{M}_{out}$ / $\dot{M}_{acc}$ in \para is now well within the limit of magneto-centrifugal jet launching models. The difference in the estimate of this ratio, as compared to results reported in \cite{Bacciotti11} is mainly due to the reduction of the value of $\dot{M}_{out}$ (by one order of magnitude),
when it is calculated with the correct formula for the population
of the atomic levels at the diagnosed physical conditions,
rather than with the approximate formula using the critical
density at an assumed temperature of 8000 K. This result has important consequences for studies aimed at constraining this ratio in the substellar mass regimes. This is discussed further in Section 5.1 below.

\subsection{Other jet emission lines}
The spectrum of the \eso jet contains a number of lines which have not been investigated as jet tracers as often as some of the traditional and well-studied jet tracers, such as the [NII]$\lambda$6583 or [SII]$\lambda$6731 lines for example. These lines warrant extra discussion. \\

{\noindent \bf HeI 1.083~$\mu$m:} The He I 1.083~$\mu$m line traces high excitation regions (excitation energy of He I at 1.083~$\mu$m is 20~eV) and temperatures in excess of $\sim$ 15,000~K. This line has been well studied in TTSs and has been found to have contributions from both the accretion onto the star \citep{Fischer08} and inner disk winds \citep{Edwards06}. Confirmation that He I 1.083~$\mu$m emission traces disk winds came from the study of the line profiles i.e. the detection of PC profiles and from spectro-astrometric studies. For example \cite{Azevedo07} use spectro-astrometry to detect an extended disk wind in the jet-less CTTS TW Hya. Observations of the He I 1.083 $\mu$m line in jets have been investigated far less frequently. \cite{Takami03} present PV diagrams of the He I 1.083 $\mu$m  emission in the DG Tau jet. They argue that shock heating must be the dominant heating mechanism in the jet in order to explain the presence of such a highly excited line. \cite{Ellerbroek12} also detect He I 1.083 $\mu$m emission in the HH~1042 and HH~1043 jets. Thus \eso is only one of a few jets where He I 1.083~$\mu$m has been identified to date. We would expect that future \xsh observations of jets would increase the number of jets for which this line is observed. The PV diagram of the \eso He I 1.083 $\mu$m emission line region is presented in Figure 1. Both knots A1 and A are detected while B and E are not seen (see discussion above). In Figure 5 we show the He I 1.083~$\mu$m emission line profile and the emission region extracted from the 2D \xsh spectrum. The part of the line tracing the jet is narrow and has a velocity of $\sim$ 0~\km\ (as seen for the other jet lines). An IPC profile which can reveal important information about the accretion is also detected. For example, \cite{Fischer08} discuss how red-shifted absorption features in the  He I 1.083 $\mu$m line can reveal information about the accretion geometry of YSOs. Specifically 
 they probed the geometry of magnetospheric accretion in CTTSs by modelling the red-shifted absorption at the He I 1.083 $\mu$m line via scattering of the stellar and veiling continua and found that the red-shifted absorption feature is sensitive to both the size of the magnetosphere of the star and the filling factor of the accretion shock. Finally some blue-shifted emission extending to $\sim$ 200~\km\ is also observed. See section 4.2.2 for a discussion of the origin of this emission.\\

{\noindent \bf [NeIII]$\lambda$3869:} The [NeIII]$\lambda$3869 line is observed in the spectrum of \eso in the region of knot A1 (see Figure 2). As the ionisation potential of Ne is high, high gas excitation
is required. For example Ne II to Ne III has an ionisation potential of 41.1~eV. Neon has been found to be excited in young stars in both high velocity shocks generated by outflows \citep{vanboekel09} and photo-evaporative disk winds \citep{Pasucci09}.  In the sample of \cite{Pasucci09} transition disk objects are compared with younger more active systems and the authors conclude that the [Ne II] emission in the transition objects comes from the photo-evaporative wind while for the less evolved objects it is more likely to be generated by outflows.  The shape and radial velocity of the Neon line can also help to distinguish between the two scenarios. If it is generated through outflow activity the line will be shifted with respect to the stellar rest velocity and will be comparable to FEL regions. If it has its origin in a photo-evaporative wind the line will be centred on the stellar rest velocity and likely asymmetric \citep{Bald12}. From Figure 1 it is clear that the most likely origin of the [NeIII]$\lambda$3869 line is in the \eso jet. The line is blue-shifted to $\sim$ 20~\km\ and is extended to 2~\arcsec. The radial velocity of this line is different from A1 as seen in the other jet tracers shown in Figure 1, in that it is not symmetric about 0~\km\ and is very definitely blue-shifted. This velocity of the  [NeIII]$\lambda$3869 line is comparable to the blue-shifted wing seen in the UVES spectra and discussed in Section. 4.2.1. \\

{\noindent \bf Refractory species:} By studying the abundances of various refractory species in jets, the depletion onto dust grains and therefore the amount of dust present in the jets can be estimated \citep{Podio06}. The amount of dust reflects the strength of the outflow shocks as the shocks will act to destroy the dust grains. \eso has a particularly rich Fe spectrum described in \cite{teresa13}. In addition, to the frequently investigated refractory species of Fe and Ca, C and Ni are also detected  in the \eso jet (see Figure 2). Both knots A1 and A are detected in the [CI]$\lambda\lambda$9824, 9850 lines. The presence of strong [C I] lines implies low excitation conditions and therefore the non-detection of [C I] in B and E supports the results of section 4.5, which show that B and E have a higher excitation that A1 and A \citep{Nisini05}. Ni+ and Fe+ have similar ionisation potentials (7.9 and 7.6 eV, respectively) and critical densities and therefore they are expected to co-exist. From Figure 2 it is seen that the [Ni II]$\lambda$7378 is only found in knot A1. \cite{Bautista96} argue that strong [Ni II] emission has been explained by the contribution of fluorescent excitation in a low density environment and that the ratio between the [Ni II]$\lambda$7378 and [Ni II]$\lambda$7412 lines can be used to distinguish between shock and fluorescent excitation. \cite{Lucy95} show that in the case of fluorescence this ratio is $\sim$ 4 but for shock excitation it is $\sim$ 10. For \eso [Ni II]$\lambda$7378 / [Ni II]$\lambda$7412 = 13 ruling out fluorescence.

\section{Discussion}

\subsection{Constraining $\dot{M}_{out}$ / $\dot{M}_{acc}$ in BDs and VLMSs}



The current sample of BDs and VLMSs with outflows is small and consists of $\sim$ 10 objects \citep{Joergens13}, ranging in mass from 0.024~\Msun\ to 0.18~\Msun\ \citep{Whelan12, Joergens12, Joergens12b}. 
Small spatial scales and the faintness of the jet emission limit the capability of studying these 
outflows with respect to the case of jets from CTTSs \citep{Ray07}, however $\dot{M}_{out}$ / $\dot{M}_{acc}$ has been investigated in some cases. 
The first investigations seemed to suggest that as mass decreases from from CTTSs to BDs, $\dot{M}_{out}$ / $\dot{M}_{acc}$ may in fact increase from the 1~$\%$ to 10~$\%$ typically measured for CTTSs. If this trend is confirmed it would present 
a challenge for current magneto-centrifugal jet launching models.
For example, \cite{Ferreira06} give an upper limit of 0.3 on the (one-sided) mass ejection to accretion ratio that can be sustained by a disk wind model and \cite{Cabrit09} discuss how high values of $\dot{M}_{out}$ / $\dot{M}_{acc}$ are energetically challenging to stellar and x-wind models. Results from the first studies of $\dot{M}_{out}$ / $\dot{M}_{acc}$ in BDs / VLMSs include, \cite{Whelan09} where this ratio is estimated for the BDs ISO-Cha~I 217, LS-RCr~A1 and ISO-Oph 102. For all three objects  $\dot{M}_{out}$ / $\dot{M}_{acc}$ (one-sided) was measured to be $\sim$ 1. More recently,  \cite{Stelzer13} report $\dot{M}_{out}$ / $\dot{M}_{acc}$ for the blue-shifted outflow of the 0.050~\Msun\ BD FU Tau~A to be $\sim$ 0.3, \cite{Joergens12b} measure $\dot{M}_{out}$ / $\dot{M}_{acc}$ to be between 1~$\%$ to 20~$\%$ for the VLMS ISO~143 and in \cite{Bacciotti11} we place this ratio for \para in the range 0.3-0.5.  It should be considered however that these investigations of accretion-ejection connection in BDs have had several limitations. 
  


Firstly, BD jet velocities are not well constrained and estimates of the full jet velocity or at least of the tangential velocities are needed to calculate $\dot{M}_{out}$. Determination of the full jet velocity from the radial velocity is hampered by the uncertainty on the inclination angle of the system. Proper motion measurements are more accurate but they have been conducted only for the Par Lup 3-4 case. In other cases 
jet velocities have been inferred from the kinematics and shape of FEL profiles \citep{Whelan09}. 
A second source of uncertainty is the estimate of n$_{e}$, normally derived from the [SII] lines. 
However, for BDs the [SII] lines have not been detected in majority of cases.
A third factor is the reliability of estimates of $\dot{M}_{acc}$ which were often only derived using the H$\alpha$ line. Additionally, the effect of extinction on the source (and thus estimate of $\dot{M}_{acc}$) and on the jet was not well known in most cases. 
Finally, and most significantly, methods for calculating $\dot{M}_{out}$ relied on uncertain values of the critical densities of the different jet tracers.

As studies of  $\dot{M}_{out}$ / $\dot{M}_{acc}$ in BDs are an important basis for comparing outflow activity in BDs and low mass stars it is essential that the difficulties outlined above are resolved. The revised analysis of $\dot{M}_{out}$ / $\dot{M}_{acc}$ in \para presented in the paper demonstrates how this can be done and hence this work is highly relevant to future studies of  $\dot{M}_{out}$ / $\dot{M}_{acc}$ in BDs and VLMSs. Indeed, for \para $\dot{M}_{out}$ / $\dot{M}_{acc}$ (one-sided) is reduced from $\sim$ 0.25 to 0.05 as compared to results presented in \cite{Bacciotti11}.  As the velocities of the \para jets are well known from proper motion studies and as 
\para contains sufficiently bright FELs for the jet parameters to be constrained \para was an ideal source for constraining  $\dot{M}_{out}$ / $\dot{M}_{acc}$. Our analysis demonstrates the importance of understanding the effects of extinction and outlines an improved method for doing this. Furthermore, the importance of estimating $\dot{M}_{acc}$ from a range of tracers is clearly established along with the fact \xsh is currently one of the best instruments for doing this. Finally, the issue with uncertainties in critical density estimates is overcome through the use of the same exact calculation for $\dot{M}_{out}$ that is usually
adopted for CTTSs jets. It is shown that the use of approximate formulas involving critical densities
can lead to severe overestimates in $\dot{M}_{out}$ for T$_{e}$ $>$
8-9 ~10$^3$ K, thus introducing a  strong  bias and it is likely that this factor has lead to substantial over-estimations of $\dot{M}_{out}$ / $\dot{M}_{acc}$ in early studies.  It would be very beneficial to now revisit the BDs studied in \cite{Whelan09} for example and investigate $\dot{M}_{out}$ / $\dot{M}_{acc}$  in the same manner as done here for Par-Lup3-4. Also note that studies aimed at measuring the proper motions and consequently velocities of known BD jets are currently under-way \citep{Whelan13}.

\subsection{Origin of \eso jet asymmetries}
\cite{Hirth94} first began the discussion of asymmetries in protostellar jets. In their paper they presented the particular cases of the asymmetric jets driven by the CTTSs RW Aur and DO Tau. They also noted, that a literature search of protostellar jets known at the time, showed that in 50~$\%$ of these jets the blue and red-shifted lobes were asymmetric in velocity. As well as velocity asymmetries the blue and red lobes of protostellar jets can also differ in their morphology and in the number of distinct knots (\eso), in the electron densities of the lobes \citep{Caratti13, Podio10} and in $\dot{M}_{out}$ \citep{Whelan09}. While no dedicated observational study of asymmetric jets has been conducted since \cite{Hirth94}, numerous examples have been observed individually using state-of-the-art observing techniques \citep{Dougados00, Melnikov09, Podio10, Caratti13}. This includes the asymmetric jet from the BD candidate ISO-ChaI 217 \citep{Whelan09, Joergens12a}. For the case of ISO-ChaI 217 the radial velocity of the red-shifted lobe was observed to be up to twice that of the blue-shifted lobe, however the exact magnitude of the velocity asymmetry will not be known until spectra taken along the derived jet PA become available. The red-shifted lobe was also found to be much brighter than the blue and the mass flux in the red flow was estimated at twice that of the blue flow. This is an interesting result as it highlights that the mechanism responsible for such asymmetries also operates at sub stellar masses. Observationally it seems that protostellar jets are more typically like \eso and symmetric jets like \para are observed less often.

In particular, jets which exhibit strong morphological symmetry, i.e. in the number and spacing of their knots are rare \citep{Zinnecker98} and even in cases where there are no velocity asymmetries it is normal that both lobes have different numbers of knots and that the spacing between these knots be variable. This is true for \para where the jets are symmetric in velocity but a counterpart to the red-shifted knot HH~600 is not seen in the blue-shifted jet. 
Morphological asymmetries can be due to non-uniformities in the ambient medium and / or variability in the frequency at which material is ejected into the different lobes. A variable frequency and velocity of ejection (as discussed for \eso in Section 4.2.1), differences in densities and in mass flux, can all be explained in terms of current jet models \citep{Fendt13, Matsakos12}. \cite{Matsakos12} used numerical simulations to investigate the possibility that asymmetric jet velocities could be introduced either due to unaligned magnetic fields or when both lobes experienced different outer pressures. That is the cause is either intrinsic to the jet launching mechanism or extrinsic and originates due to inhomogenities in the ambient medium. Overall they found that both multi-polar magnetic moments and non-uniform environments could equally well explain the observed asymmetries. The idea of an inhomogeneous environment causing velocity asymmetries has been used before to explain cases of asymmetric jets where $\dot{M}_{out}$ is not found to be different in the two lobes \citep{Podio10, Melnikov09}. \cite{Fendt13} also explore numerically methods for generating jet asymmetries which are intrinsic to the launch mechanism. To do this they begin with a highly symmetric jet and then disturb the symmetry in the disk to induce asymmetries in the jets. Interestingly they find that the disk asymmetries result in outflows where $\dot{M}_{out}$ can differ by up to 20$\%$ in the two lobes. Comparing $\dot{M}_{out}$ in the blue-shifted lobe of \eso with $\dot{M}_{out}$ in the red-shifted lobe, it is seen that $\dot{M}_{out}$ red is $\sim$ 60$\%$ of $\dot{M}_{out}$ blue. Therefore this case would fit in well with the models of \citep{Fendt13}.  $\dot{M}_{out}$ blue is the average of $\dot{M}_{out}$ measured for knots A1, A and B while $\dot{M}_{out}$ red is the value measured for knot E


\section{Summary}

{In this paper the accretion and outflow properties of \eso and Par-Lup3-4, derived from analysis of their \xsh spectra, are described. This analysis is a continuation of the preliminary study described in 
\cite{Bacciotti11}. Here the full set of data 
in all the different portions of the outflows, obtained after careful revision of  the flux calibration and the effects of reddening, are presented. The application of improved diagnostics are described and, in particular, new and more accurate estimates of  $\dot{M}_{out}$ / $\dot{M}_{acc}$
to be compared with predictions of models of the jet launching are offered. 
\eso represents an extreme example of a YSO with an edge-on accretion disk and a very rich spectrum 
of lines from high ionisation species. Thus, the treatment of this object presented here will be important for future studies of similar systems. 
\para is one of a small sample of BDs and VLMSs known to drive jets. 
As \para is the only object in this group where $\dot{M}_{out}$ / $\dot{M}_{acc}$ has been estimated without large observational uncertainties, it is singularly important in the context of studies of jet launching at BD masses.
Our  main findings can be summarised as follows: }

\begin{itemize}

\item{The spectrum of \eso contains an abundance of jet lines but weak accretion tracers, and some lines like Br$\gamma$ are not present. It is hypothesised that the weakness of the accretion lines is due to the obscuration of the accretion zone by its edge-on accretion disk. This has been hypothesised for other YSOs whose spectra are rich in outflow lines but show comparatively weak accretion lines \citep{Looper10a}. In comparison \para has all the well-studied accretion lines (e.g. CaII triplet, Br$\gamma$) but a much smaller number of jet lines, of a lower excitation level. \para is also postulated to have an edge-on disk but the comparison with \eso in terms of accretion tracers points to its disk having a lower inclination. 
Tables of all the emission lines detected in both sources are given in the Appendix. }

\item{A review of the kinematical properties of the \eso jet is presented in Section 4.2. The low inclination of the jet means that what can be understood about the jet close to the driving source is limited. However, asymmetries are detected in the jet which can be explained in terms of differences in ejection rates between the two lobes. The \eso \Ha\ line region is interesting in that it has a prominent blue-wing in the \xsh spectrum and a stronger red-shifted wing in the UVES spectrum. The \Ha\ line profile is divided into a narrow and a broad component. The narrow component is made-up of emission from the jet while the origin of the broad component is likely made up of emission from both infall and outflow. The \para jet is far more symmetric than the \eso jet. }

\item{The extinction is an important quantity as it strongly affects measurements of  $\dot{M}_{acc}$.
By comparing the spectra of \eso and \para with \xsh spectra of zero extinction Class III templates, a 
relative  extinction of 1.5~mag is estimated for ESO-H$\alpha$ 574 while no relative extinction 
correction is needed for Par-Lup3-4. The fact that the relative extinction-corrected luminosity of both sources 
is still considerably less than the average luminosity of other similar YSOs in their parent cloud
suggests that they are affected by grey extinction. This is due to obscuration by their disks and it is estimated that their luminosities are reduced by factors of 150 and 25 for \eso and \para, respectively.}
\item{
Measurements of $\dot{M}_{acc}$ before and after obscuration correction are shown presented.
The obscuration corrected values of log($\dot{M}_{acc}$) are -9.15 $\pm$ 0.45 \,\Msun yr$^{-1}$ and -9.30 $\pm$ 0.27 \,\Msun yr$^{-1}$ for \eso and Par-Lup3-4, respectively. 
By also estimating $\dot{M}_{acc}$ from the indirect indicator [OI]$\lambda$6300
{\it emitted from an extended region,} 
the obscuration factors can be tested. As $\dot{M}_{acc}$ from [OI] agrees with $\dot{M}_{acc}$ 
made from the direct accretion tracers and corrected for obscuration, the estimate of $\dot{M}_{acc}$ in \para is likely accurate. For \eso it is found that the obscuration correction does not completely solve the uncertainty in  $\dot{M}_{acc}$. It is argued that veiling and scattering effects also impact on the extinction measurement. }

\item{In Section 4.5 the Balmer decrements for the knots A1 and A in the \eso flow, and for the \para source spectrum are fitted with Case B and LTE optically thick and thin models. For A1 and A results are consistent with an origin in the jet. The spatial range of A1 includes emission from \eso itself and this result further emphasises that the bulk of the emission in A1 comes from the outflow. For \para the analysis shows that the decrements are best fitted with a Case B model with a density and temperature of 10$^{10}$~cm$^{-3}$ and 10,000~K. This is compatible with densities and temperatures derived for a magnetospheric accretion flow.}

\item{Several line ratios are used to probe the ionisation, excitation, and temperature in knots A1, A, B, and E, and in the \para red and blue jets.  
It is found that the \eso jet is characterised by temperatures around 
1.4 $\times$10$^4$ K and a hydrogen ionisation fraction $>$ 0.65, with knots B and E bearing 
the higher excitation. The electron density decreases from 4000~cm$^{-3}$ to 150~cm$^{-3}$ from inner to outer knots. 
In contrast, the \para jet has a low degree of ionisation (x$_{e}$ = 2-6 10$^{-3}$),
but a higher temperature (2.5 - 3 $\times$ 10$^4$ K) and electron density (4 $\times$ 10$^3$ cm$^{-3}$).
These results explain why the \eso has a spectrum much richer in lines from high ionised species than the \para spectrum. 
The values of parameters found on-source are within the range of values derived by Giannini et al. (2013)
from the analysis of the iron lines.

It is argued that no high excitation lines are detected in knots B and E of ESO-H$\alpha$ 574, despite their high excitation, as the electron density is too low in these knots. 
B and E have a higher shock velocity that than either A1, A or both lobes of the \para flow. This conclusion is borne out through comparison with the models of \cite{Hartigan94}. }

\item { 
The improved diagnostic analysis and the adoption of an exact calculation based on a 5-level model atom  
allowed for an improved estimate (as compared to Bacciotti et al. 2011) of the mass outflow rate $\dot{M}_{out}$ 
in the jets. When $\dot{M}_{out}$ is combined with an updated estimate considering grey extinction of $\dot{M}_{acc}$, the estimate of the ratio $\dot{M}_{out}$ / $\dot{M}_{acc}$ is also improved. For ESO-H$\alpha$ 574, $\dot{M}_{out}$ / $\dot{M}_{acc}$ in the red and blue flows is 0.5 (+1.0)(-0.2) and 0.3 (+0.6)(-0.1), respectively. For Par-Lup3-4, $\dot{M}_{out}$ / $\dot{M}_{acc}$ is 0.05 (+0.10)(-0.02) in both the red and blue flows. 
The improved estimate for \para $\dot{M}_{out}$ / $\dot{M}_{acc}$ is now within the range predicted by current jet launching models. The \eso estimate still has large errors and it is not excluded that it lies within the limits predicted by models. 
Various biases which have likely affected previous studies of $\dot{M}_{out}$ / $\dot{M}_{acc}$ in BDs and VLMSs are discussed. The analysis of \para offers an important example of how estimates of $\dot{M}_{out}$ / $\dot{M}_{acc}$ in the lowest mass objects can be greatly improved by overcoming such biases. It should be noted however, that further improvements are needed, beyond the scope of this paper, to ensure that all biases have been overcome, e.g. the effects of scattering and veiling.}

\end{itemize}

\acknowledgements{E.T. Whelan acknowledges financial support from the BMWi/DLR grant FKZ 50 OR 1309 and the. L.P. acknowledges the funding from the FP7 Intra-European Marie Curie Fellowship (PIEF-GA-2009-253896). We thank G. Cupani, V. D'Odorico, P. Goldoni and A. Modigliani
 for their help with the X-Shooter pipeline.  F. Getman and G. Capasso
 are acknowledgement for the installation of the different  pipeline
 versions at Capodimonte.
 We also thank the ESO staff, in particular F. Patat for suggestions
 in OB preparation and C. Martayan for support during the observations.
 Financial support from INAF is also acknowledged.}

{}


\clearpage \onecolumn

\begin{appendix}

\section{Tables of identified lines and line fluxes}

\begin{longtable}{rp{1.5cm}p{2.5cm}rp{0.5cm}rp{3.cm}} 
\caption{ESO-H$\alpha$ 574 emission line fluxes. Fluxes refer to knot A1, are in units of ergs/s/cm$^{2}$ $\times$ 10$^{-17}$ and are not extinction corrected.  Where the identified lines are also found in knots A, B or E the observed wavelengths and fluxes are also given. The regions over which the spectra for A1, A, B and E are extracted are shown in Figure 2. The horizontal lines separate the UVB, VIS and NIR arms. \tablefoottext{a}{blended lines.}}\\
\hline \hline 
\multicolumn{2}{c}{\hspace{2cm} Identification}\\
\cline{1-4}
$\lambda_{air}$ (\AA)  &Ion  &Type &E$_{u}$ (cm$^{-1}$)  &Knot &$\lambda_{obs}$ (\AA)  &Flux ( $\times$ 10$^{-17}$ ergs/s/cm$^{2}$)  
 \\  
 \hline
3721.9$^a$	&	H I	   &	 14	- 2                                         &	109119.2	&		&	3721.8	&	3.5 $\pm$ 0.5
\\												
3721.6$^a$	&	[S III]	   &	$^1$S$_0$-$^3$P$_1$	                       &	27161.00	         &		&		         &	
\\												
3726.0	&	[O II]	   &	$^2$D$_{3/2}$-$^4$S$_{3/2}$                   &	26830.6   	&		&	3726.0	&	76.8 $\pm$ 0.5
\\												
	         &		   &		                                                             &	                    	&	A	&	3726.0	&	71.7 $\pm$ 0.5
\\												
	         &		   &		                                                             &		                  & 	B	&	3726.0	&	18.3 $\pm$ 0.5
\\												
	         &		   &		                                                             &		                 &	E	&	3726.6	&	9.2 $\pm$ 0.5
\\												
3728.8	&	[O II]	   &	$^2$D$_{5/2}$-$^4$S$_{3/2}$               &	26810.6	        &		&	3728.8	&	38.0 $\pm$ 0.5
\\												
	         &		   &		                                                             &		                  &	A	&	3728.8	&	46.7 $\pm$ 0.5
\\												
	         &		   &		                                                             &		                 &	B	&	3728.8	&	22.5 $\pm$ 0.5
\\												
	         &		   &		                                                             &		                 &	E	&	3729.0	&	13.3 $\pm$ 0.5
\\												
3734.4	&	H I	   &	 13	- 2                                                              &	109029.8       &		&	3734.3	&	1.3 $\pm$ 0.5
\\												
3750.2	&	H I	   &	 12	- 2                                                   &	108917.1       &		&	3750.3	&	1.4 $\pm$ 0.5
\\
3769.9$^a$	& [Co II]	   &	$a^3$H$_{5}$-$a^3$F$_{3}$	              &	27469.1   	&		&		         &	1.7 $\pm$ 0.5
\\													
3770.2$^a$	&	Ti I 	   &	$s^3$F$_{4}$-$a^3$G$_{3}$	              &	41624.2	        &		&		         &	
\\																							
3770.6	&	H I	   &	11 - 2	                                                   &	108772.3	&		&	3771.3	&	1.7 $\pm$ 0.5
\\												
3797.9	&	H I	   &	10 - 2	                                                  &	108582.0	&		&	3797.8	&	2.3 $\pm$ 0.5
\\												
3835.3	&	H I	    &	9 - 2 	                                                           &	108324.7	&		&	3835.3	&	4.9 $\pm$ 0.5
\\												
3869.1	&     [Ne III]    &	$^1$D$_{2}$-$^3$P$_{2}$	            &	25838.7	         &		&	3968.9      &	11.5 $\pm$ 0.5
\\												
3889.0	&	H I	    &	8 - 2 	                                                           &	107965.1	&		&	3888.9	&	1.2 $\pm$ 0.5
\\												
3933.7	&	Ca II 	    &	$^2$P$_{3/2}$-$^2$S$_{1/2}$	            &	25414.4	          &		&	3933.7	&	16.6 $\pm$ 0.3
\\												
3967.8	&	[Ne III]  &	$^1$D$_{2}$-$^3$P$_{1}$	            &	25838.7	          &		&	3967.4	&	4.7 $\pm$ 0.3
\\												
3968.2$^a$	&	[Fe II]   &	$a^{4}$G$_{5/2}$-$a^{6}$D$_{3/2}$   &	26055.4	          &		&	3968.5	&	12.5 $\pm$ 0.3
\\
3968.2$^a$	&	[Ti II]	   &	$c^2$D$_{5/2}$-$a^4$F$_{3/2}$	    &	25192.8    	&		&		          &	
\\												
3968.4$^a$	&	Ca II 	&	$^2$P$_{1/2}$-$^2$S$_{1/2}$              &	25191.5	          &		&		          &	
\\												
3970.1	&	H I	&	 7 - 2	                                                             &	107440.5	 &		&	3970.1	 &	10.0 $\pm$ 0.3
\\												
	          &		&		                                                             &		                    &	A       &	3970.1	 &	3.35 $\pm$ 0.3
\\												
4068.6	&	[S II]	&	$^2$P$_{3/2}$-$^4$S$_{3/2}$               &	24571.5	            &		&	4068.6	  &	149.0 $\pm$ 0.4
\\												
	          &		&	           	                                                    &	                    	   &	A	&	4068.5	&	41.7 $\pm$ 0.4
\\												
4076.3	&	[S II]	&	$^2$P$_{1/2}$-$^4$S$_{3/2}$	               &	24524.8	            &		&	4076.3	&	50.1 $\pm$ 0.4
\\												
	         &		&	           	                                                     &	                    	   &	A	&	4076.3	&	15.9 $\pm$ 0.4
\\												
4101.7	&	H I	&	6 - 2	                                                              &	106632.2	   &		&	4101.8	&	13.4 $\pm$ 0.4
\\												
	          &		&		                                                              &		                      &	A	&	4101.8	&	8.8 $\pm$ 0.4
\\												
4114.5	&	[Fe II]	&	$b^2$H$_{11/2}$-$a^4$F$_{9/2}$	&	26170.2	&		&	4114.5	&	2.7 $\pm$ 0.4
\\												
4211.1	&	[Fe II]	&	$b^2$H$_{11/2}$-$a^4$F$_{7/2}$	&	26170.2	&		&	4211.1	&	1.2 $\pm$ 0.4
\\												
4244.0	&	[Fe II]	&	$a^4$G$_{11/2}$-$a^4$F$_{9/2}$	&	25428.8	&		&	4244.0	&	6.0 $\pm$ 0.5
\\												
4276.8	&	[Fe II]	&	$a^4$G$_{9/2}$-$a^4$F$_{7/2}$	&	25805.3	&		&	4276.9	&	2.5 $\pm$ 0.5
\\																							
4287.4	&	[Fe II]	&	$a^6$S$_{5/2}$-$a^6$D$_{9/2}$	&	23317.6	&		&	4287.4	&	6.6 $\pm$ 0.4
\\												
4319.6	&	[Fe II]	&	$a^4$G$_{7/2}$-$a^4$F$_{5/2}$	&	25981.6	&		&	4319.6	&	1.9 $\pm$ 0.3
\\												
4340.5	&	H I	         &	 5 - 2	                                                        &105291.7	&		&	4340.4	&	25.0 $\pm$ 0.4
\\												
	         &		         &		                                                        &                            &	A	&	4340.5	&	18 $\pm$ 0.4
\\												
4346.9	&	[Fe II]	&	$a^4$G$_{11/2}$-$a^4$F$_{7/2}$	&	25428.8	&		&	4346.6	&	0.7 $\pm$ 0.3
\\												
4347.4	&	[Fe II]	&	$b^4$D$_{5/2}$-$a^4$D$_{5/2}$	&	31387.9	&		&	4347.6	&	1.0 $\pm$ 0.3
\\												
4352.8	&	[Fe II]	&	$a^4$G$_{9/2}$-$a^4$F$_{5/2}$	&	25805.3	&		&	4352.8	&	1.0 $\pm$ 0.3
\\												
4359.3	&	[Fe II]	&	$a^6$S$_{5/2}$-$a^6$D$_{7/2}$	&	23317.6	&		&	4359.3	&	4.8 $\pm$ 0.3
\\												
4362.7	&	C I	&	$^3$D$_{3}$-$^3$D$_{3}$	&	87002.3	&		&	4362.5	&	0.8 $\pm$ 0.3
\\												
4363.2	&	[O III]	&	$^1$S$_{0}$-$^1$D$_{2}$	&	43185.7	&		&	4363.4	&	3.0 $\pm$ 0.3
\\												
4413.8	&	[Fe II]	&	$a^6$S$_{5/2}$-$a^6$D$_{5/2}$	&	23317.6	&		&	4413.8	&	3.1 $\pm$ 0.3
\\												
4416.3	&	[Fe II]	&	$b^4$F$_{9/2}$-$a^6$D$_{9/2}$	&	22637.2	&		&	4416.3	&	4.4 $\pm$ 0.3
\\												
4432.4	&	[Fe II]	&	$b^4$F$_{5/2}$-$a^6$D$_{7/2}$	&	22939.4	&		&	4432.7	&	1.6 $\pm$ 0.3
\\												
4438.9	&	[Fe II]	&	$b^4$D$_{1/2}$-$a^4$D$_{1/2}$	&	31368.5	&		&	4439.2	&	1.0 $\pm$ 0.4
\\												
4452.1	&	[Fe II]	&	$a^6$S$_{5/2}$-$a^6$D$_{3/2}$	&	23317.6	&		&	4452.2	&	2.8 $\pm$ 0.4
\\												
4457.9	&	[Fe II]	&	$b^4$F$_{7/2}$-$a^6$D$_{7/2}$	&	22810.4	&		&	4457.9	&	2.3 $\pm$ 0.3
\\												
4471.7	&	He I 	&	$^3$D$_{1}$-$^3$P$_{0}$	&	191444.6	&		&	4471.5	&	4.0 $\pm$ 0.4
\\												
4571.1$^a$	&	K I	&	$^2$D$_{3/2}$-$^2$P$_{3/2}$	&	34913.3	&	&4571.5	&		3.2 $\pm$ 0.4	
\\												
4571.1$^a$	&	Mg I]	&	$^3$P$_{1}$-$^1$S$_{0}$	&	21870.5	&		&		&	
\\												
4580.8	&	[Cr II]	&	$a^4$P$_{3/2}$-$a^6$S$_{5/2}$	&	21824.1	&		&	4580.8	&	1.4 $\pm$ 0.3
\\												
4607.0	&	[Fe III]	&	$^3$F$_{3}$-$^5$D$_{4}$	&	21699.9	&		&	4606.8	&	1.0 $\pm$ 0.3
\\												
4658.1	&	[Fe III]	&	$^3$F$_{4}$-$^5$D$_{4}$	&	21462.2	&		&	4658.1	&	9.2 $\pm$ 0.3
\\												
4685.7	&	He II 	&	4 - 3	&	411478.0	&		&	4685.7	&	2.1 $\pm$ 0.3
\\												
4701.5	&	[Fe III]	&	$^3$F$_{3}$-$^5$D$_{3}$	&	21699.9	&		&	4701.6	&	3.4 $\pm$ 0.2
\\												
4713.4	&	He I 	&	$^3$S$_{1}$-$^3$P$_{0}$	&	190298.2	&		&	4713.4	&	1.0 $\pm$ 0.2
\\												
4728.1	&	[Fe II]	&	$b^4$P$_{3/2}$-$a^6$D$_{5/2}$	&	21812.1	&		&	4728.1	&	1.0 $\pm$ 0.3
\\												
4733.0$^a$	&	Fe II]	         &	$^6$D$_{5/2}$-$y^4$D$_{5/2}$	&	83812.3	&		&	4733.0	&	0.7 $\pm$ 0.3
\\												
4733.0$^a$	&	Fe III]	&	$^3$F$_{2}$-$^1$F$_{3}$	&	118163.6	&		&		&	
\\												
4733.9	&	[Fe III]	&	$^3$F$_{2}$-$^5$D$_{2}$	&	21857.2	&		&	4734.1	&	1.5 $\pm$ 0.3
\\												
4754.7	&	[Fe III]	&	$^3$F$_{4}$-$^5$D$_{3}$	&	21462.2	&		&	4755.0	&	2.2 $\pm$ 0.4
\\												
4769.4	&	[Fe III]	&	$^3$F$_{3}$-$^5$D$_{2}$	&	21699.9	&		&	4769.6	&	1.8 $\pm$ 0.3
\\												
4774.7	&	[Fe II]	&	$b^4$F$_{7/2}$-$a^4$F$_{9/2}$	&	22810.4	&		&	4774.4	&	1.1 $\pm$ 0.3
\\												
4777.7$^a$	&	[Fe III]	&	$^3$F$_{2}$-$^5$D$_{1}$	&	21857.2	&		&	4777.7	&	1.0 $\pm$ 0.3
\\												
4778.0$^a$	&	[Fe II]	&	$a^2$D${3/2}$-$a^6$D$_{7/2}$	&	21308.0	&		&			
\\												
4798.3	&	[Fe II]	&	$b^4$P$_{3/2}$-$a^6$D$_{1/2}$	&	21812.1	&		&	4798.0	&	0.7 $\pm$ 0.3
\\												
4814.5	&	[Fe II]	&	$b^4$F$_{9/2}$-$a^4$F$_{9/2}$	&	22637.2	&		&	4814.5	&	3.7 $\pm$ 0.4
\\												
4861.3	&	H I	&	4 - 2	&	102823.9	&		&	4861.3	&	63.2 $\pm$ 0.3
\\												
	&		&		&		&	A	&	4861.3	&	46.7 $\pm$ 0.3
\\												
	&		&		&		&	B	&	4861.4	&	10.8 $\pm$ 0.3
\\												
	&		&		&		&	E	&	4861.8	&	16.3 $\pm$ 0.3
\\												
4881.0	&	[Fe III] 	&	$^3$H$_{4}$-$^5$D$_{4}$	&	40481.9	&		&	4881.1	&4.6 $\pm$ 0.4	
\\												
4889.6$^a$	&	[Fe II]	&	$b^4$P$_{5/2}$-$a^6$D$_{7/2}$	&	20830.6	&		&	4889.6	&	3.8 $\pm$ 0.3
\\												
4889.7$^a$	&	[Fe II]	&	$a^2$D$_{3/2}$-$a^6$D$_{3/2}$	&	21308.0	&		&		&	
\\												
4905.3	&	[Fe II]	&	$b^4$F$_{7/2}$-$a^4$F$_{7/2}$	&	22810.4	&		&	4905.4	&	2.0 $\pm$ 0.3
\\												
4930.5	&	[Fe III]	&	$^3$P$_{0}$-$^5$D$_{1}$	&	21208.5	&		&	4930.5	&	1.2 $\pm$ 0.2
\\												
4931.2	&	[O III]	&	$^1$D$_{2}$-$^3$P$_{0}$	&	20273.3	&		&	4930.8	&	0.9 $\pm$ 0.2
\\												
4950.7	&	[Fe II]	&	$b^4$F$_{3/2}$-$a^4$F$_{5/2}$	&	23031.3	&		&	4950.7	&	0.9 $\pm$ 0.2
\\												
4958.9	&	[O III]	&	$^1$D$_{2}$-$^3$P$_{1}$	&	20273.3	&		&	4958.9	&	11.6 $\pm$ 0.2
\\												
	&		&		&	                            	&	A	&	4958.8	&	4.4 $\pm$ 0.2
\\												
4973.4	&	[Fe II]	&	$b^4$F$_{5/2}$-$a^4$F$_{5/2}$	&	22939.4	&		&	4973.4	&	0.9 $\pm$ 0.2
\\												
4987.2	&	[Fe III]	&	$^3$H$_{4}$-$^5$D$_{3}$	&	20481.9	&		&	4987.4	&	1.1 $\pm$ 0.2
\\												
5006.8	&	[O III]	&	$^1$D$_{2}$-$^3$P$_{2}$	&	20273.3	&		&	5006.8	&	34.5 $\pm$ 0.3
\\												
	&		&		&		&	A	&	5006.7	&	14.7 $\pm$ 0.3
\\												
5011.3	&	[Fe III]	&	$^3$P$_{1}$-$^5$D$_{2}$	&	20688.4	&		&	5011.2	&	2.2 $\pm$ 0.3
\\												
5015.7	&	He I 	&	$^1$P$_{1}$-$^1$S$_{0}$	&	186209.5	&		&	5015.8	&	1.3 $\pm$ 0.3
\\												
5111.6	&	[Fe II]	&	$a^4$H$_{11/2}$-$a^4$F$_{9/2}$	&	21430.4	&		&	5111.5	&	2.1 $\pm$ 0.3
\\												
5158.8	&	[Fe II]	&	$a^4$H$_{13/2}$-$a^4$F$_{9/2}$	&	21251.6	&		&	5158.7	&	15.7 $\pm$ 0.3
\\												
5164.0$^a$	&	[Fe II]	&	$a^2$F$_{7/2}$-$a^4$D$_{7/2}$	&	27314.9	&		&	5164.4	&	2.0 $\pm$ 0.3
\\												
5164.5$^a$	&	[Cr II]	&	$c^2$F$_{7/2}$-$a^4$G$_{9/2}$	&	39877.1	&		&		&	
\\												
5197.9	&	[N I]	&	$^2$D$_{3/2}$-$^4$S$_{3/2}$	&	19233.2	&		&	5197.9	&	9.7 $\pm$ 0.3
\\												
	&		&		&		&	A	&	5197.9	&	4.5 $\pm$ 0.3
\\												
5200.3	&	[N I]	&	$^2$D$_{5/2}$-$^4$S$_{3/2}$	&	19224.5	&		&	5200.3	&	6.4 $\pm$ 0.3
\\												
	&		&		&		&	A	&	5200.3	&	3.2 $\pm$ 0.3
\\												
5220.1	&	[Fe II]	&	$a^4$H$_{9/2}$-$a^4$F$_{7/2}$	&	21581.6	&		&	5220.0	&	2.1 $\pm$ 0.4
\\												
5261.6	&	[Fe II]	&	$a^4$H$_{11/2}$-$a^4$F$_{7/2}$	&	21430.4	&		&	5261.6	&	7.0 $\pm$ 0.3
\\												
5270.4	&	[Fe III]	&	$^3$P$_{2}$-$^5$D$_{3}$	&	19404.8	&		&	5270.5	&	4.0 $\pm$ 0.3
\\												
5273.3	&	[Fe II]	&	$b^4$P$_{5/2}$-$a^4$F$_{9/2}$	&	20830.6	&		&	5273.3	&	3.1 $\pm$ 0.3
\\												
5333.6	&	[Fe II]	&	$a^4$H$_{9/2}$-$a^4$F$_{5/2}$	&	21581.6	&		&	5333.6	&	3.6 $\pm$ 0.3
\\												
5337.2	&	[Cr II]	&	$c^4$D$_{3/2}$-$a^4$D$_{3/2}$	&	38362.4	&		&	5336.8	&	3.4 $\pm$ 0.3
\\												
5376.5	&	[Fe II]	&	$a^4$H$_{7/2}$-$a^4$F$_{3/2}$	&	21711.9	&		&	5376.4	&	3.0 $\pm$ 0.3
\\												
5412.0$^a$	&	[Fe III]	&	$^3$P$_{2}$-$^5$D$_{1}$	&	19404.8	&		&	5412.1	&	1.0 $\pm$ 0.3
\\												
5412.7$^a$	&	[Fe II]	&	$a^2$D$_{3/2}$-$a^4$F$_{5/2}$	&	21308.0	&		&		&	
\\												
5413.3	&	[Fe II]	&	$a^2$H$_{11/2}$-$a^4$F$_{9/2}$	&	20340.3	&		&	5413.1	&	1.2 $\pm$ 0.3
\\												
5527.3$^a$	&	[Fe II]	&	$a^2$D$_{5/2}$-$a^4$F$_{7/2}$	&	20517.0	&		&	5527.2	&	4.3 $\pm$ 0.5
\\												
5527.6$^a$	&	[Fe II]	&	$b^2$P$_{1/2}$-$a^4$D$_{1/2}$	&	26932.8	&		&		&	
\\ \\
\hline
\\									
5754.6	&	[N II]	&	$^1$S$1_{0}$-$^1$D$_{2}$	&	32688.6	&		&	5754.4	& 	5.9 $\pm$ 0.3
\\  
5875.9	&	He I 	&	$^3$D$_{1}$-$^3$P$_{0}$	&	186101.7		&	      &	5875.9	&	8.1 $\pm$ 0.3
\\												
5890.1	&	Cr II]	&	$^6$D$_{3/2}$-$v^4$F$_{5/2}$	&	107414.7	 &		&	5890.2	&	5.3 $\pm$ 0.3
\\												
5896.0	&	Fe I	&	$i^5$D$_{2}$-$x^5$F$_{2}$	&	57974.1		&	&	5896.0	&	2.3 $\pm$ 0.3
\\												
5980.1	&	Fe I] 	&	$x^1$F$_{3}$-$d^3$F$_{4}$	&	53763.3		&	&	5980.1	&	1.9 $\pm$ 0.3
\\												
6300.3	&	[O I]	&	$^1$D$_{1}$-$^3$P$_{2}$	&	15867.9		&	&	6300.5	&	116.0 $\pm$ 0.2
\\												
	         &		&		                                               &			&A	         &	6300.5	&	56.3 $\pm$ 0.2
\\												
6312.1	&	[S III]	&	$^1$S$_{0}$-$^1$D$_{2}$	&	27161.0	 &		&	6312.3	&	4.7 $\pm$ 0.2
\\												
6363.7	&	[O I]	&	$^1$D$_{2}$-$^3$P$_{1}$	&	15867.9	&		&	6363.9	&	37.2 $\pm$ 0.2
\\												
	         &		&		                                               &			 &A	          &	6364.0	&	14.3 $\pm$ 0.2
\\												
6548.0	&	[N II]	&	$^1$D$_{2}$-$^3$P$_{1}$	&	15316.2	 &		&	6548.2	&	43.6 $\pm$ 0.2
\\												
	&		&		&			&A	&	6548.2	&	28.0 $\pm$ 0.2
\\												
	&		&		&			&B	&	6548.1	&	4.1 $\pm$ 0.2
\\												
6562.8	&	H I	&	3 - 2	&	97492.3	&		&	6562.9	&	248.1 $\pm$ 0.2
\\												
	&		&		&			&A	&	6562.9	&	189.0 $\pm$ 0.2
\\												
	&		&		&			&B	&	6562.8	&	41.2 $\pm$ 0.2
\\												
	&		&		&			&E	&	6563.4	&	58.9 $\pm$ 0.2
\\												
6583.5	&	[N II]	&	$^1$D$_{2}$-$^3$P$_{2}$	&	15316.2	&		&	6583.6	&	126.2 $\pm$ 0.2
\\												
	&		&		&			&A	&	6583.6	&	83.9 $\pm$ 0.2
\\												
	&		&		&			&B	&	6583.4	&	11.2 $\pm$ 0.2
\\												
	&		&		&			&E	&	6583.4	&	6.0 $\pm$ 0.2
\\												
6678.2	&	He I 	&	$^1$D$_{2}$-$^1$P$_{1}$	&	186105.1	&		&	6678.3	&	1.5 $\pm$ 0.3
\\												
6716.4	&	[S II]	&	$^2$D$_{5/2}$-$^4$S$_{3/2}$	&	14884.7	&		&	6716.6	&	206.9 $\pm$ 0.2
\\												
	&		&		&			&A	&	6716.6	&	116.0 $\pm$ 0.2
\\												
	&		&		&			&B	&	6716.4	&	26.7 $\pm$ 0.2
\\												
	&		&		&			&E	&	6716.9	&	6.3 $\pm$ 0.2
\\												
6730.8	&	[S II]	&	$^2$D$_{3/2}$-$^4$S$_{3/2}$	&	14852.9		&	&	6731.0	&	352.5 $\pm$ 0.2
\\												
	&		&		&			&A	&	6731.0	&	162.0 $\pm$ 0.2
\\												
	&		&		&			&B	&	6730.8	&	22.1 $\pm$ 0.2
\\												
	&		&		&			&E	&	6731.4	&	5.0 $\pm$ 0.2
\\												
7065.7	&	He I 	&	$^3$S$_{1}$-$^3$P$_{0}$	&	183236.9		&	&	7065.4	&	0.8 $\pm$ 0.2
\\												
7135.8	&	[Ar III]	&	$^1$D$_{2}$-$^3$P$_{2}$	&	14010.0	&		&	7136.0	&	3.5 $\pm$ 0.2
\\												
7155.2	&	[Fe II]	&	$a^2$G$_{9/2}$-$a^4$F$_{9/2}$	&	15844.7	&		&	7155.3	&	26.9 $\pm$ 0.2
\\												
7172.0	&	[Fe II]	&	$a^2$G$_{7/2}$-$a^4$F$_{7/2}$	&	16369.4	&		&	7172.1	&	7.4 $\pm$ 0.2
\\												
7291.5	&	[Ca II]	&	$^2$D$_{5/2}$-$^2$S$_{1/2}$	&	13710.9	&		&	7291.6	&	20.2 $\pm$ 0.2
\\												
7319.9	&	[O II]	&	$^2$D$_{3/2}$-$^2$P$_{5/2}$	 &	40468.0		&	&	7320.1	&	45.1 $\pm$ 0.2
\\												
	&		&		&			&A	&	7320.0	&	13.4 $\pm$ 0.2
\\												
7323.8	&	[Ca II]	&	$^2$D$_{3/2}$-$^2$S$_{1/2}$	&	13650.2		&	&	7324.1	&	12.9 $\pm$ 0.2
\\												
7330.7	&	[O II]	&	$^2$P$_{3/2}$-$^2$D$_{3/2}$	&	40468.0		&	&	7330.4	&	35.2 $\pm$ 0.2
\\												
7377.8	&	[Ni II]	&	$^2$F$_{7/2}$-$^2$D$_{5/2}$	&	13550.4		&	&	7378.1	&	11.4 $\pm$ 0.2
\\												
7388.2	&	[Fe II]	&	$a^2$G$_{7/2}$-$a^4$F$_{5/2}$	&	16369.4		&	&	7388.4	&	4.3 $\pm$ 0.2
\\												
7411.6	&	[Ni II]   	&	$^2$F$_{7/2}$-$^2$D$_{3/2}$		&	14955.6		&	&	7412.1	&	0.9 $\pm$ 0.2
\\												
7452.5	&	[Fe II]	&	$a^2$G$_{9/2}$-$a^4$F$_{7/2}$	&	15844.7	&		&	7452.7	&	8.0 $\pm$ 0.2
\\			
7637.5	&	[Fe II]	&	$a^4$P$_{5/2}$-$a^6$D$_{7/2}$	&	13474.4	&		&	7637.6	&	4.3 $\pm$ 0.2
\\									
7686.9	&	[Fe II]	&	$a^4$P$_{3/2}$-$a^6$D$_{5/2}$	&	13673.2		&	&	7687.1	&	2.2 $\pm$ 0.2
\\												
7703.6$^a$	&	[Mn II]	&	$a^5$G$_{5}$-$a^5$D$_{3}$	&	27571.3		&	&	7703.8	&	7.3 $\pm$ 0.3
\\												
7703.8$^a$	&	[Co II]	&	$c^3$F$_{3}$-$a^3$H$_{4}$	&	40879.4	&		&		&	
\\												
8000.1	&	[Cr II]	&	$a^6$D$_{9/2}$-$a^6$S$_{5/2}$	&	12496.4		&	&	8000.3	&	3.2 $\pm$ 0.2
\\												
8125.3	&	[Cr II]	&	$a^6$D$_{7/2}$-$a^6$S$_{5/2}$	&	12303.9		&	&	8125.5	&	3.2 $\pm$ 0.2
\\												
8229.6	&	[Cr II]	&	$a^6$D$_{5/2}$-$a^6$S$_{5/2}$	&	12147.8		&	&	8229.9	&	2.9 $\pm$ 0.2
\\												
8308.5	&	[Cr II]	&	$a^6$D$_{3/2}$-$a^6$S$_{5/2}$	&	12032.6		&	&	8308.8	&	1.7 $\pm$ 0.2
\\												
8498.0	&	Ca II 	&	$^2$P$_{3/2}$-$^2$D$_{3/2}$	&	25414.4	&		&	8498.3	&	1.9 $\pm$ 0.2
\\												
8542.1	&	Ca II 	&	$^2$P$_{3/2}$-$^2$D$_{5/2}$	&	25414.4	&		&	8452.4	&	2.2 $\pm$ 0.2
\\												
8578.7	&	[Cl II]	&	$^1$D$_{2}$-$^3$P$_{2}$	&	11653.6	&		&	8578.9	&	6.2 $\pm$ 0.2
\\												
8617.0	&	[Fe II]	&	$a^4$P$_{5/2}$-$a^4$F$_{9/2}$	&	13474.4	&		&	8617.2	&	2.8 $\pm$ 0.2
\\												
8662.1	&	Ca II 	&	$^2$P$_{1/2}$-$^2$D$_{3/2}$	&	25191.5	&		&	8662.4	&	1.2 $\pm$ 0.2
\\												
8727.1	&	[C I]	&	$^1$S$_{0}$-$^1$D$_{2}$	&	21648.0	&		&	8727.3	&	2.3 $\pm$ 0.3
\\												
8891.9	&	[Fe II]	&	$a^4$P$_{3/2}$-$a^4$F$_{7/2}$	&	13673.2	&		&	8892.2	&	10.5 $\pm$ 0.2
\\												
9033.5	&	[Fe II]	&	$a^4$P$_{1/2}$-$a^4$F$_{5/2}$	&	13904.8		&	&	9033.7	&	4.2 $\pm$ 0.4
\\												
9051.9	&	[Fe II]	&	$a^4$P$_{5/2}$-$a^4$F$_{7/2}$	&	13474.4	&		&	9052.2	&	7.8 $\pm$ 0.2
\\												
9068.6	&	[S III]	&	$^1$D$_{2}$-$^3$P$_{1}$	&	11322.7		&	&	9069.2	&	21.6 $\pm$ 0.3
\\												
9123.6	&	[Cl II]	&	$^1$D$_{2}$-$^3$P$_{1}$	&	11653.6		&	&	9123.7	&	2.2 $\pm$ 0.3
\\												
9226.6	&	[Fe II]	&	$a^4$P$_{3/2}$-$a^4$F$_{5/2}$	&	13673.2	&		&	9226.8	&	7.4 $\pm$ 0.2
\\												
9267.6	&	[Fe II]	&	$a^4$P$_{1/2}$-$a^4$F$_{3/2}$	&	13904.8	&		&	9267.7	&	5.6 $\pm$ 0.3
\\												
9530.6	&	[S III]	&	$^1$D$_{2}$-$^3$P$_{2}$	&	11322.7		&	&	9531.2	&	64.9 $\pm$ 0.3
\\												
	&		&		&			&A	&	9530.9	&	16.4 $\pm$ 0.3
\\												
9824.1	&	[C I]	&	$^1$D$_{2}$-$^3$P$_{1}$	&	10192.6	&		&	9824.5	&	4.0 $\pm$ 0.3
\\												
9850.3	&	[C I]	&	$^1$D$_{2}$-$^3$P$_{2}$	&	10192.6	&		&	9850.6	&	11.4 $\pm$ 0.3
\\ \\																				
  \hline
 \\  
12086.7	    &	[S II]	      &	$^2$P$_{3/2}$-$^2$D$_{3/2}$	      &	24571.5    	&		&	10286.5	&	50 $\pm$ 1
\\												
	             &		     &		                                                     &		                   &A		&	10286.6	&	20 $\pm$ 1
\\												
10320.5	    &	[S II]	     &	$^2$P$_{3/2}$-$^2$D$_{5/2}$       &	24571.5   	&		&	10320.4	&	64 $\pm$ 1
\\												
	              &		     &		                                                      &		          &A		&	10320.4	&	20 $\pm$ 1
\\												
10336.4	    &	[S II]	     &	$^2$P$_{1/2}$-$^2$D$_{3/2}$	       &	24524.8	&		&	10336.3	&	50 $\pm$ 1
\\												
	              &		     &		                                                      &		          &A		&	10336.3	&	14 $\pm$ 1
\\												
10370.5	     &	[S II]	     &	$^2$P$_{1/2}$-$^2$D$_{5/2}$	       &	24524.8	&		&	10370.4	&	22 $\pm$ 1
\\												
	              &		     &		                                                       &		          &A		&	10370.4	&	7 $\pm$ 1
\\												
10397.7$^a$	    &	[N I]	     &	$^2$P$_{3/2}$-$^2$D$_{5/2}$	       &	28839.3	&		&	10398.0	&	4 $\pm$1
\\												
10398.2$^a$	    &	[N I]	    &	$^2$P$_{1/2}$-$^2$D$_{5/2}$	       &	28838.9	 & 		&	               	&	
\\												
10407.2$^a$	    &	[N I]	    &	$^2$P$_{3/2}$-$^2$D$_{3/2}$	       &	28839.3	 &		&	10407.2	&	3 $\pm$ 1
\\												
10407.6$^a$	    &	[N I]	    &	$^2$P$_{1/2}$-$^2$D$_{3/2}$	       &	28838.9	 &		&		          &	
\\												
10830.3	&	He I 	&	$^3$P$_{1}$-$^3$S$_{1}$	        &	169087.0 &		&	10830.1	&	132.7 $\pm$ 0.8
\\												
	           &		&		                                                       &		           &A		&	10830.3	&	56.2 $\pm$ 0.3
\\						
10938.0	&	H I	&	0 - 3 	&		106632.2	&	&	10937.8	&	14.5 $\pm$ 0.8
\\												
11618.9	&	[Ti II]	&	$a^2$G$_{7/2}$-$a^4$F$_{9/2}$	&	8997.7	&		&	11618.9	&	6 $\pm$ 1
\\												
12485.4	&	[Fe II]	&	$a^4$D$_{5/2}$-$a^6$D$_{7/2}$	&	8391.9	&		&	12484.9	&	2.3 $\pm$ 0.4
\\												
12566.8	&	[Fe II]	&	$a^4$D$_{7/2}$-$a^6$D$_{9/2}$	&	7955.3	&		&	12566.6	&	82.7 $\pm$ 0.7
\\												
	&		&		&		&A		&	12566.4	&	16.7 $\pm$ 0.2
\\												
12610.9	&	[Cr I]	&	$a^5$D$_{2}$-$a^7$S$_{3}$	&	7927.5	&		&	12611.0	&	3 $\pm$ 1
\\												
12787.8	&	[Fe II]	&	$a^4$D$_{3/2}$-$a^6$D$_{3/2}$	&	8680.45	&		&	12787.5	&	11 $\pm$ 2
\\												
12818.1	&	H I	&	 5 - 3 	&	105291.7	&		&	12817.7	&	13 $\pm$ 1
\\												
12942.7	&	[Fe II]	&	$a^4$D$_{5/2}$-$a^6$D$_{5/2}$	&	8391.9	&		&	12942.5	&	13.8 $\pm$ 0.8
\\												
12977.7	&	[Fe II]	&	$a^4$D$_{3/2}$-$a^6$D$_{1/2}$	&	8680.4	&		&	12977.8	&	3.7 $\pm$ 0.8
\\												
13205.5	&	[Fe II]	&	$a^4$D$_{7/2}$-$a^6$D$_{7/2}$	&	7955.30	&		&	13205.3	&	26.9 $\pm$ 0.8
\\												
13277.8	&	[Fe II]	&	$a^4$D$_{5/2}$-$a^6$D$_{3/2}$	&	8391.9	&		&	13277.8	&	9.5 $\pm$ 0.9
\\												
15334.7	&	[Fe II]	&	$a^4$D$_{5/2}$-$a^4$F$_{9/2}$	&	8391.9	&		&	15334.4	&	13.9 $\pm$ 0.6
\\												
15994.7	&	[Fe II]	&	$a^4$D$_{3/2}$-$a^4$F$_{7/2}$	&	8680.5	&		&	15994.7	&	14.4 $\pm$ 0.5
\\												
16435.5	&	[Fe II]	&	$a^4$D$_{7/2}$-$a^4$F$_{9/2}$	&	7955.3	&		&	16435.3	&	72.5 $\pm$ 0.7
\\												
16637.7	&	[Fe II]	&	$a^4$D$_{1/2}$-$a^4$F$_{5/2}$	&	8846.8	&		&	16637.4	&	7.2 $\pm$ 0.6
\\												
16768.8	&	[Fe II]	&	$a^4$D$_{5/2}$-$a^4$F$_{7/2}$	&	8391.9	&		&	16768.1	&	16.7 $\pm$ 0.6
\\												
17475.0	&	H$_{2}$ 	&	1-0S(7)	&		&		&	1.74746	&	9.3 $\pm$ 0.9
\\												
17971.0	&	[Fe II]	&	$a^4$D$_{3/2}$-$a^4$F$_{3/2}$	&	8680.5	&		&	17970.7	&	4 $\pm$ 1
\\												
18093.9	&	[Fe II]	&	$a^4$D$_{7/2}$-$a^4$F$_{7/2}$	&	7955.3	&		&	18093.5	&	16 $\pm$ 2
\\												
19570.2  &	H$_{2}$ 	&	1-0S(3)	&		&		&	19570.8	&	69.7 $\pm$ 0.2
\\												
21212.5	&	H$_{2}$ 	&	1-0S(1)	&		&		&	21212.5	&	39.2 $\pm$ 0.2
\\												
22226.8	&	H$_{2}$ 	&	1-0S(0)	&		&		&	22227.0	&	8.4 $\pm$ 0.3
\\																							
22471.0	&	H$_{2}$ 	&	2-1S(1)	&		&		&	22470.0	&	4.8 $\pm$ 0.3
\\																				
\hline\hline
\end{longtable}

\newpage

\begin{longtable}{rp{1.5cm}p{2.5cm}rp{0.5cm}rp{3.cm}}
\caption{Par-Lup3-4 emission line fluxes. Fluxes refer to the source spectrum (extracted over the range -0.5~\arcsec to 0.5~\arcsec), are in units of ergs/s/cm$^{2}$ $\times$ 10$^{-17}$ and are not extinction corrected. Where the identified lines are also found in the extended blue (B) and red-shifted (R) jets the observed wavelengths and fluxes are also given. The spatial regions corresponding to B and R are marked in Figure 7. The horizontal lines separate the UVB, VIS and NIR arms. \tablefoottext{a}{blended lines.}}\\
\hline \hline 
\multicolumn{2}{c}{\hspace{2cm} Identification}\\
\cline{1-4}
$\lambda_{air}$ ($\AA$)  &Ion  &Type &E$_{u}$ (cm$^{-1}$)  &Jet &$\lambda_{obs}$ ($\AA$)  &Flux ($\times$ 10$^{-17}$ ergs/s/cm$^{2}$ )  
 \\  
  \hline
3770.6	&	H I	&	11 - 2  &	108772.34	&		&3769.9		&	2.5 $\pm$ 0.4
\\												
3797.9	&	H I	&  10 - 2&	108581.99	&		&	3797.2	&	7.1 $\pm$ 0.3
\\												
3835.3	&	H I	&	9 - 2	&	108324.73	&		&	3834.9	&	9.0 $\pm$ 0.4
\\												
3889.0	&	H I	&	8 - 2	&	107965.06	&		&	3888.2	&	10.5 $\pm$ 0.3
\\												
3933.7	&	Ca II	&	$^2$P$_{3/2}$-$^2$S$_{1/2}$	&	25414.40	&		&	3933.2	&	28.2 $\pm$ 0.3
\\												
3968.5	&	Ca II\tablefootmark{a}	&	$^2$P$_{1/2}$-$^2$S$_{1/2}$	&	25191.51	&		&	3968.1	&	27.6 $\pm$ 0.2
\\
3970.1	&	H I\tablefootmark{a}	&	7 - 2	                                                             &	107440.45	 &		&	 &	
\\													
4068.6	&	[SII]	&	$^2$P$_{3/2}$-$^4$S$_{3/2}$	&	24571.54	&		&	4068.1	&	98.2 $\pm$ 0.2
\\												
4076.3	&	[SII]	&	$^2$P$_{1/2}$-$^4$S$_{3/2}$	&	24524.83	&		&	4075.9	&	27.8 $\pm$ 0.2
\\												
4101.7	&	H I	& 6	- 2 &	106632.17	&		&	4101.0	&	15.6 $\pm$ 0.2
\\												
4114.5	&	[Fe II]	&	$b^2$H$_{11/2}$-$a^4$F$_{9/2}$	&	26170.18	&		&	4114.0	&	1.0 $\pm$ 0.4
\\												
4244.0	&	[Fe II]	&	$a^4$G$_{11/2}$-$a^4$F$_{9/2}$	&	25428.78	&		&	4243.4	&	3.1 $\pm$ 0.2
\\												
4276.8	&	[Fe II]	&	$a^4$G$_{9/2}$-$a^4$F$_{7/2}$	&	25805.33	&		&	4276.3	&	1.7 $\pm$ 0.2
\\												
4319.6	&	[Fe II]	&	$a^4$G$_{7/2}$-$a^4$F$_{5/2}$	&	25981.63	&		&	4319.1	&	1.3 $\pm$ 0.2
\\												
4340.5	&	H I	          &	 5  - 2                                                   	&	105291.66	&		&	4339.7	&	25.0 $\pm$ 0.2
\\												
4346.9	&	[Fe II]	&	$a^4$G$_{11/2}$-$a^4$F$_{7/2}$	&	25428.78	&		&	4346.1	&	1.0 $\pm$ 0.2
\\												
4352.8	&	[Fe II]	&	$a^4$G$_{9/2}$-$a^4$F$_{5/2}$	&	25805.33	&		&	4351.5	&	1.9 $\pm$ 0.2
\\												
4359.3	&	[Fe II]	&	$a^6$S$_{5/2}$-$a^6$D$_{7/2}$	&	23317.63	&		&	4358.6	&	2.7 $\pm$ 0.2
\\												
4413.8	&	[Fe II]	&	$a^6$S$_{5/2}$-$a^6$D$_{5/2}$	&	23317.63	&		&	4413.3	&	2.2 $\pm$ 0.2
\\												
4416.3	&	[Fe II]	&	$b^4$F$_{9/2}$-$a^6$D$_{9/2}$	&	22637.21	&		&	4415.7	&	3.0 $\pm$ 0.2
\\												
4452.1	&	[Fe II]	&	$a^6$S$_{5/2}$-$a^6$D$_{3/2}$	&	23317.63	&		&	4451.4	&	1.0 $\pm$ 0.2
\\												
4457.9	&	[Fe II]	&	$b^4$F$_{7/2}$-$a^6$D$_{7/2}$	&	22810.36	&		&	4457.4	&	1.4 $\pm$ 0.2
\\												
4571.1	&	Mg I]	&	$^3$P$_{1}$-$^1$S$_{0}$	&	21870.46	&		&	4570.6	&	6.4 $\pm$ 0.2
\\												
4728.1	&	[Fe II]	&	$b^4$P$_{3/2}$-$a^6$D$_{5/2}$	&	21812.06	&		&	4727.5	&	1.2 $\pm$ 0.1
\\												
4814.5	&	[Fe II]	&	$b^4$F$_{9/2}$-$a^4$F$_{9/2}$	&	22637.21	&		&	4814.0	&	2.0 $\pm$ 0.1
\\												
4861.3	&	H I	&	 4 - 2	&	102823.90	&		&	4860.6	&	60.4 $\pm$ 0.2
\\												
4889.6$^a$	&	[Fe II]	&	$b^4$P$_{5/2}$-$a^6$D$_{7/2}$	&	20830.58	&		&	4888.9	&	1.6 $\pm$ 0.1
\\											
4889.7$^a$	&	[Fe II]	&	$a^2$D$_{3/2}$-$a^6$D$_{3/2}$	&	21308.0	&		&		&	
\\												
4905.3	&	[Fe II]	&	$b^4$F$_{7/2}$-$a^4$F$_{7/2}$	&	22810.36	&		&	4905.0	&	1.3 $\pm$ 0.1
\\												
5015.7	&	He I	&	$^1$P$_{1}$-$^1$S$_{0}$	&	186209.47	&		&	5015.1	&	5.8 $\pm$ 0.2
\\												
5111.6	&	[Fe II]	&	$a^4$H$_{11/2}$-$a^4$F$_{9/2}$	&	21430.36	&		&	5110.9	&	1.7 $\pm$ 0.2
\\												
5158.8	&	[Fe II]	&	$a^4$H$_{13/2}$-$a^4$F$_{9/2}$	&	21251.61	&		&	5158.0	&	6.5 $\pm$ 0.2
\\												
5197.9	&	[N I]	&	$^2$D$_{3/2}$-$^4$S$_{3/2}$	&	19233.18	&		&	5197.5	&	6.1 $\pm$ 0.2
\\												
5200.2	&	[N I]	&	$^2$D$_{5/2}$-$^4$S$_{3/2}$	&	19224.46	&		&	5199.8	&	3.4 $\pm$ 0.2
\\												
5220.1	&	[Fe II]	&	$a^4$H$_{9/2}$-$^a4$F$_{7/2}$	&	21581.64	&		&	5219.5	&	1.4 $\pm$ 0.2
\\												
5261.6	&	[Fe II]	&	$a^4$H$_{11/2}$-$a^4$F$_{7/2}$	&	21430.36	&		&	5260.9	&	3.6 $\pm$ 0.2
\\												
5273.3	&	[Fe II]	&	$b^4$P$_{5/2}$-$a^4$F$_{9/2}$	&	20830.58	&		&	5272.7	&	1.5 $\pm$ 0.2
\\												
5333.6	&	[Fe II]	&	$a^4$H$_{9/2}$-$a^4$F$_{5/2}$	&	21581.64	&		&	5333.0	&	2.8 $\pm$ 0.2
\\												
5376.5	&	[Fe II]	&	$a^4$H$_{7/2}$-$a^4$F$_{3/2}$	&	21711.92	&		&	5375.8	&	2.4 $\pm$ 0.2
\\												
5412.7	&	[Fe II]	&	$a^2$D$_{3/2}$-$a^4$F$_{5/2}$	&	21308.04	&		&	5411.9	&	0.8 $\pm$ 0.2
\\												
5527.3$^a$	&	[Fe II]	&	$a^2$D$_{5/2}$-$a^4$F$_{7/2}$	&	20516.96	&		&	5526.7	&	3.3 $\pm$ 0.2
\\												
5527.6$^a$	&	[Fe II]	&	$b^2$P$_{1/2}$-$a^4$D$_{1/2}$	&	26932.75	&		&		&	
\\												
5577.3	&	[O I]	&	$^1$S$_{0}$-$^1$S$_{2}$	&	33792.58	&		&	5576.7	&	11.6 $\pm$ 0.3
\\ \\																				
\hline
 \\  
5875.9	&	He I	&	$^3$D$_{1}$-$^3$P$_{0}$	&	186101.7	&		&	5875.3	&	8.2 $\pm$ 0.2
\\												
5890.1	&	Cr II]	&	$^6$D$_{3/2}$-$v^4$F$_{5/2}$	&	107414.7	&		&	5889.2	&	7.6 $\pm$ 0.2
\\												
5895.9	&	Fe I	&	$i^5$D$_{2}$-$x^5$F$_{2}$	&	57974.1	&		&	5895.2	&	5.3 $\pm$ 0.2
\\												
6300.3	&	[O I]	&	$^1$D$_{1}$-$^3$P$_{2}$	&	15867.9	&		&	6299.7	&	248.3 $\pm$ 0.3
\\
	&		&		&		&B		&6299.5		&54.4 $\pm$ 0.3	
\\												
	&		&		&		&R		&6299.8		&63.6 $\pm$ 0.3	
\\												
6363.7	&	[O I]	&	$^1$D$_{2}$-$^3$P$_{1}$	&	15867.9	&		&	6363.1	&	81.8 $\pm$ 0.3
\\	
	&		&		&		&B		&6362.8		&16.4 $\pm$ 0.3	
\\												
	&		&		&		&R		&6363.1		&19.2 $\pm$ 0.3		
\\																						
6548.0	&	[N II]	&	$^1$D$_{2}$-$^3$P$_{1}$	&	15316.2	&		&	6547.7	&	4.7 $\pm$ 0.2
\\																							
6562.8	&	H I	&	3 - 2	&	97492.31	&		&	6562.4	&	841.4 $\pm$ 0.2
\\																							
6583.5	&	[N II]	&	$^1$D$_{2}$-$^3$P$_{2}$	&	15316.2	&		&	6583.0	&	10.5 $\pm$ 0.2
\\
	&		&		&		&B		&6583.3		&3.8 $\pm$ 0.2	
\\												
	&		&		&		&R		&6582.6		&3.5 $\pm$ 0.2	
\\																							
6678.2	&	He I	&	$^1$D$_{2}$-$^1$P$_{1}$	&	186105.1	&		&	6677.7	&	3.6 $\pm$ 0.2
\\												
6716.4	&	[S II]	&	$^2$D$_{5/2}$-$^4$S$_{3/2}$	&	14884.7	&		&	6716.0	&	29.6 $\pm$ 0.2
\\												
	&		&		&		&B		&	6715.6	&	10.1 $\pm$ 0.2
\\												
	&		&		&		&R		&	6716.4	&	14.1 $\pm$ 0.2
\\												
6730.8	&	[S II]	&	$^2$D$_{3/2}$-$^4$S$_{3/2}$	&	14852.9	&		&	6730.4	&	56.3 $\pm$ 0.2
\\												
	&		&		&		&B		&	6729.9	&	15.1 $\pm$ 0.2
\\												
	&		&		&		&R		&	6730.8	&	22.7 $\pm$ 0.2
\\												
7065.7	&	He I	&	$^3$S$_{1}$-$^3$P$_{0}$	&	183236.9	&		&	7064.8	&	2.6 $\pm$ 0.3
\\												
7155.2	&	[Fe II]	&	$a^2$G$_{9/2}$-$a^4$F$_{9/2}$	&	15844.7	&		&	7154.5	&	9.5 $\pm$ 0.1
\\																							
7172.0	&	[Fe II]	&	$a^2$G$_{7/2}$-$a^4$F$_{7/2}$	&	16369.4	&		&	7171.3	&	2.9 $\pm$ 0.1
\\												
7291.4	&	[Ca II]	&	$^2$D$_{5/2}$-$^2$S$_{1/2}$	&	13710.9	&		&	7291.1	&	2.1 $\pm$ 0.2
\\												
7319.9	&	[O II]	&	$^2$P$_{5/2}$-$^2$D$_{3/2}$	&	40468.01	&		&	7319.2	&	2.9 $\pm$ 0.2
\\												
7330.7	&	[O II]	&	$^2$P$_{3/2}$-$^2$D$_{3/2}$	&	40468.01	&		&	7329.7	&	1.7 $\pm$ 0.2
\\												
7377.8	&	[Ni II]	&	$^2$F$_{7/2}$-$^2$D$_{5/2}$	&	13550.4	&		&	7377.2	&	3.4 $\pm$ 0.2
\\												
7388.2	&	[Fe II]	&	$a^2$G$_{7/2}$-$a^4$F$_{5/2}$	&	16369.4	&		&	7387.6	&	3.1 $\pm$ 0.2
\\												
7411.6	&	[Ni II]	  &$^2$F$_{5/2}$-$^2$D$_{3/2}$		&	14955.0	&		&	7412.1	&0.9 $\pm$ 0.2	
\\												
7452.5	&	[Fe II]	&	$a^2$G$_{9/2}$-$a^4$F$_{7/2}$	&	15844.7	&		&	7451.8	&	3.3 $\pm$ 0.2
\\												
8498.0	&	Ca II	&	$^2$P$_{3/2}$-$^2$D$_{3/2}$	&	25414.4	&		&	8497.3	&	36.8 $\pm$ 0.2
\\												
8542.1	&	Ca II	&	$^2$P$_{3/2}$-$^2$D$_{5/2}$	&	25414.4	&		&	8541.2	&	39.6 $\pm$ 0.2
\\												
8617.0	&	[Fe II]	&	$a^4$P$_{5/2}$-$a^4$F$_{9/2}$	&	13474.4	&		&	8616.1	&	6.8 $\pm$ 0.3
\\												
8662.1	&	Ca II	&	$^2$P$_{1/2}$-$^2$D$_{3/2}$	&	25191.5	&		&	8661.3	&	35.5 $\pm$ 0.2
\\												
8727.1	&	[C I]	&	$^1$S$_{0}$-$^1$D$_{2}$	&	21648.0	&		&	8726.6	&	36.1 $\pm$ 0.2
\\												
8891.9	&	[Fe II]	&	$a^4$P$_{3/2}$-$a^4$F$_{7/2}$	&	2430.1	&		&	8891.2	&	2.9 $\pm$ 0.8
\\												
9226.6$^a$	&	[Fe II]	&	$a^4$P$_{3/2}$-$a^4$F$_{3/2}$	&	13673.2	&		&	9227.4	&	20.8 $\pm$ 0.8
\\
9227.8$^a$       &    He I  & 	$^3$P-$^3$D                   & 	196935.4    	&
\\  										
9850.3	&	[C I]	&	$^1$D$_{2}$-$^3$P$_{2}$			&10192.6	 &	&	9849.6	&	85.4 $\pm$ 0.2
\\ \\																				
  \hline
\\    
10286.7	&	[S II]	&	$^2$P$_{3/2}$-$^2$D$_{3/2}$	&	24571.5	&		&	10285.4	&	38.8 $\pm$ 0.6
\\												
10320.5	&	[S II]	&	$^2$P$_{3/2}$-$^2$D$_{5/2}$	&	24571.5	&		&	10319.2	&	48.3 $\pm$ 0.6
\\												
10336.4	&	[S II]	&	$^2$P$_{1/2}$-$^2$D$_{3/2}$	&	24524.8	&		&	10335.1	&	30.2 $\pm$ 0.6
\\												
10370.5	&	[S II]	&	$^2$P$_{1/2}$-$^2$D$_{5/2}$	&	24524.8	&		&	10369.3	&	14.6 $\pm$ 0.6
\\												
10397.7$^a$	&	[N I]	&	$^2$P$_{3/2}$-$^2$D$_{5/2}$	&	28839.3	&		&	10396.5	&	40.0 $\pm$ 0.6
\\												
10398.2$^a$	&	[N I]	&	$^2$P$_{1/2}$-$^2$D$_{5/2}$	&	28839.9	&		&		&	
\\												
10407.2$^a$	&	[N I]	&	$^2$P$_{3/2}$-$^2$D$_{3/2}$	&	28839.3	&		&	10406.1	&	31.2 $\pm$ 0.6
\\												
10407.6$^a$	&	[N I]	&	$^2$P$_{1/2}$-$^2$D$_{3/2}$	&	28838.9	&		&		&
\\												
10830.3	&	He I	&	$^3$P$_{1}$-$^3$S$_{1}$	&	169087.0	&		&	10828.0	&	78.0 $\pm$ 0.6
\\												
10938.0	&	H I	&	6 - 3 	&	106632.2	&		&	10936.3	&	44.2 $\pm$ 0.6
\\												
12566.8	&	[Fe II]	&	$a^4$D$_{7/2}$-$a^6$D$_{9/2}$	&	7955.3	&		&	12565.6	&	13.1 $\pm$ 0.7
\\												
12787.8	&	[Fe II]	&	$a^4$D$_{3/2}$-$a^6$D$_{3/2}$	&	8680.5	&		&	12786.1	&	3.7 $\pm$ 0.7
\\												
12818.1	&	H I	&	 5 - 3	&	105291.7	&		&	12815.5	&	73.8 $\pm$ 0.7
\\												
12942.7	&	[Fe II]	&	$a^4$D$_{5/2}$-$a^6$D$_{5/2}$	&	8391.9	&		&	12941.4	&	3.4 $\pm$ 0.7
\\												
16435.5	&	[Fe II]	&	$a^4$D$_{7/2}$-$a^4$F$_{9/2}$	&	7955.3	&		&	16433.3	&	11.2 $\pm$ 0.8
\\												
21212.5		&	H$_{2}$ &	 1-0S(1)		&		&		&	21210.0	&	11.6 $\pm$ 0.8 
\\												
21655.3	&	H I	&	7 - 4	&	107440.5	&		&	21650.4	&	16.4 $\pm$ 0.6												
\\																				
\hline\hline
\end{longtable}

\end{appendix}

\end{document}